\documentclass[tightenlines,twocolumn,amsmath,amssymb,showkeys,twoside,aps,floatfix,prb]{revtex4}

\usepackage{ifthen}
\newboolean{JoT}
\setboolean{JoT}{false}

\usepackage{amsmath} 
\usepackage{graphicx} 
\usepackage{dcolumn} 
\usepackage{bm} 
\usepackage{color} 
\usepackage{mathrsfs}
\usepackage{fancyhdr}
\usepackage{verbatim}
\usepackage[colorlinks,letterpaper,dvips]{hyperref}

\newcommand{\laur}{LA-UR 10-05227, v0.9, \emph{Accepted in Journal of Turbulence, Jan 7, 2011}}
\newcommand{\version}{\laur}

\newcommand{\ti}{Extending the Langevin model to variable-density pressure-gradient-driven turbulence}

\newcommand{\kw}{Probability density function method; Langevin equation; Variable-density turbulence; Pressure-gradient-driven flows; Small-scale anisotropy}

\hypersetup{citecolor=blue,
			linkcolor=blue,
			urlcolor=blue,
			pdftitle=\ti,
			pdfauthor=J. Bakosi and J. R. Ristorcelli,
			pdfkeywords=\kw}
\usepackage{hypernat}

\begin{document}

\newcommand{\bv}[1]{{\mbox{\boldmath$#1$}}} 
\newcommand{\fmean}[1]{{\langle{#1}\rangle}} 
\newcommand{\rmean}[1]{{\overline{#1}}} 
\newcommand{\rf}{'} 
\newcommand{\ff}{''} 
\newcommand{\rv}[1]{\rmean{#1\rf^2}} 
\newcommand{\fv}[1]{\fmean{#1\ff^2}} 
\newcommand{\rs}[1]{\rmean{#1\rf^3}} 
\newcommand{\rk}[1]{\rmean{#1\rf^4}} 
\newcommand{\ld}[1]{\frac{\mathrm{d}#1}{\mathrm{d}t}} 
\newcommand{\ild}[1]{\mathrm{d}#1/\mathrm{d}t} 
\newcommand{\sd}[1]{\frac{\mathrm{D}#1}{\mathrm{D}t}} 
\newcommand{\msd}[1]{\frac{\mathrm{\overline{D}}#1}{\mathrm{\overline{D}}t}} 
\newcommand{\erf}[1]{\mathrm{erf}\left(#1\right)} 
\newcommand{\sm}{{\scriptstyle\mathcal{M}}} 

\newcommand{\vd}{d} 
\newcommand{\vdrf}{d\rf} 
\newcommand{\ivdrf}{d\rf} 

\newcommand{\Eqr}[1]{(\ref{#1})}
\newcommand{\Eqre}[1]{Eq.~(\ref{#1})}
\newcommand{\Eqrs}[1]{(\ref{#1})}
\newcommand{\Eqres}[1]{Eqs.~(\ref{#1})}
\newcommand{\Fig}[1]{figure~\ref{#1}}
\newcommand{\Figs}[1]{figures~\ref{#1}}
\newcommand{\Fige}[1]{Figure~\ref{#1}}
\newcommand{\Figse}[1]{Figures~\ref{#1}}
\newcommand{\fig}[1]{fig.~\ref{#1}}
\newcommand{\figs}[1]{figs.~\ref{#1}}
\newcommand{\fige}[1]{Fig.~\ref{#1}}
\newcommand{\figse}[1]{Figs.~\ref{#1}}

\ifthenelse{\boolean{JoT}}
{}
{\pagestyle{fancy}
 \fancyhead{}
 \fancyhead[LE,RO]{\thepage}
 \fancyfoot{}
 \chead{\texttt{\version}}
 \renewcommand{\headrulewidth}{0pt}}

\title{\ti\vspace{0pt}\\\small\texttt{\version}}

\ifthenelse{\boolean{JoT}}
{\author{J. Bakosi and J. R. Ristorcelli$^{\ast}$\thanks{$^\ast$Email: \{jbakosi,jrrj\}@lanl.gov\vspace{6pt}}\\
\vspace{6pt}{\em{Los Alamos National Laboratory, Los Alamos, NM 87545, USA}}\\
\vspace{6pt}\received{\version}}}
{\author{J. Bakosi}\author{J. R. Ristorcelli\\\small\texttt{\{jbakosi,jrrj\}@lanl.gov}}\affiliation{Los Alamos National Laboratory, Los Alamos, NM 87545, USA}}

\ifthenelse{\boolean{JoT}}
{\maketitle}
{}

\begin{abstract}
We extend the generalized Langevin model,\cite{Haworth_86} originally developed for the Lagrangian fluid particle velocity in constant-density shear-driven turbulence, to variable-density (VD) pressure-gradient-driven flows. VD effects due to non-uniform mass concentrations (e.g.\ mixing of different species) are considered. In the extended model large density fluctuations leading to large differential fluid accelerations are accounted for. This is an essential ingredient to represent the strong coupling between the density and velocity fields in VD hydrodynamics driven by active scalar mixing. The small scale anisotropy, a fundamentally \emph{``non-Kolmogorovian''} feature of pressure-gradient-driven flows, is captured by a tensorial stochastic diffusion term. The extension is so constructed that it reduces to the original Langevin model in the limit of constant density.

We show that coupling a Lagrangian mass-density particle model to the proposed extended velocity equation results in a statistical representation of VD turbulence that has important benefits. Namely, the effects of the mass flux and the specific volume, both essential in the prediction of VD flows, are retained in closed form and require no explicit closure assumptions.

The paper seeks to describe a theoretical framework necessary for subsequent applications. We derive the rigorous mathematical consequences of assuming a particular functional form of the stochastic momentum equation coupled to the stochastic density field in variable-density flows. Our aim is to develop a joint model for variable-density pressure-gradient-driven turbulence and mixing, such as occurs due to the Rayleigh-Taylor instability. A previous article\cite{Bakosi_10} discussed VD mixing and developed a stochastic Lagrangian model equation for the mass-density. Second in the series, this article develops the momentum equation for VD hydrodynamics. A third, forthcoming paper will combine these ideas on mixing and hydrodynamics into a comprehensive framework: it will specify a joint model for the coupled problem and validate it by numerically computing joint statistics of a Rayleigh-Taylor flow at several Atwood numbers.
\end{abstract}

\ifthenelse{\boolean{JoT}}
{}
{}

\ifthenelse{\boolean{JoT}}
{}
{\maketitle}

\section{Introduction}
\citet{Haworth_86} developed the stochastic model for the Lagrangian fluid particle velocity, $v_i^*$, in constant-density shear-driven turbulence,
\begin{equation}
\mathrm{d}v_i^* = -\rmean{p},_i/\varrho_0\mathrm{d}t + G_{ij}\left(v^*_j-\rmean{v}_j\right)\mathrm{d}t + \left(C_0\varepsilon\right)^{1/2}\mathrm{d}W_i,\label{eq:orGLM}
\end{equation}
where $\varrho_0$, $p$ and $\varepsilon$ are the constant density, pressure and the dissipation rate of turbulent kinetic energy, respectively. The coefficients, $G_{ij}$ and $C_0$, are specified by the particular model of choice. \Eqre{eq:orGLM} is a stochastic differential equation (SDE) of the diffusion type, where $\mathrm{d}W_i(t)$ is an isotropic Wiener process\cite{Gardiner_09} and the overbar denotes the ensemble average.

\Eqre{eq:orGLM} is used to represent the probability density function (PDF) of velocity. In PDF methods\cite{Pope_85} a transport equation is solved for the joint PDF of several flow variables. In most cases the equation is of Fokker-Planck-type and its numerical solution amounts to following a large number of Lagrangian particles in a Monte-Carlo fashion, whose properties are governed by an equivalent system of SDEs, such as \Eqre{eq:orGLM} for the particle velocity. These computational particles do not correspond to real fluid particles, but the statistics of an ensemble are representative of the local statistical behavior of the flow over many realizations.

\subsection{Variable-density pressure-gradient-driven turbulence}
Our goal is to extend \Eqre{eq:orGLM} to \emph{variable-density} (VD) \emph{pres\-sure-gradient-driven turbulence} (PGDT), a class of flows that is fundamentally different from constant-density shear-driven flows. A classical example of PGDT is the Rayleigh-Taylor (RT) instability of an interface between two fluids of different densities in a gravitational field that points to the opposite of the density gradient.\cite{Rayleigh_82,Taylor_50,Sharp_84} RT flows are important in several geophysical, astrophysical and engineering applications.

\textbf{Boussinesq limit.} If the densities of the mixing fluids are commensurate, PGDT is treated in the Boussinesq limit: the density fluctuations are small compared to the mean density and are only important in the body force term of the Navier-Stokes equation. A discussion of RT-turbulence in the Boussinesq case, with a self-similar analysis at the second moment level, is given by Ristorcelli and Clark.\cite{Ristorcelli_Clark_04}

\textbf{Variable-density case.} In situations where the fluids have vastly different densities the term \emph{variable-density} is used to distinguish it from the Boussinesq case. A commonly used measure of VD effects is the Atwood number,
\begin{equation}
A = \frac{\varrho_2-\varrho_1}{\varrho_2+\varrho_1} \qquad \Longrightarrow \qquad \frac{\varrho_2}{\varrho_1}=\frac{1+A}{1-A},\label{eq:A}
\end{equation}
where $\varrho_1\!<\!\varrho_2$ denote the constant densities of the pure fluids in a binary mixture. $A\approx0$ corresponds to the Boussinesq case, while $A\to1$ to largely disparate species densities, $\varrho_1\ll\varrho_2$. Mixing at high Atwood numbers, compared to the Boussinesq case, is accompanied by several new effects: (1) the advection term in the Navier-Stokes equation gives rise to cubic non-linearities, (2) additional non-linearities in the molecular diffusion terms, (3) a dynamic mean pressure gradient, and (4) the asymmetry of the mixing layer. In these flows the mean specific volume becomes an important independent variable. As a consequence of these new phenomena both hydrodynamics and mixing exhibit qualitatively different behavior compared to the Boussinesq case.\cite{Livescu_07,Livescu_08,Livescu_09,Livescu_09b,Livescu_09c}

\textbf{Challenges associated with VD-PGDT.} The above description highlights the tight coupling of the density and velocity fields in VD flows. This poses unique challenges, compared to classical (and more extensively researched) constant-density shear-driven turbulence. They can be enumerated as follows:
\begin{enumerate}
\item \emph{Mixing-driven:} The hydrodynamics of VD-PGDT may primarily be driven by material mixing. Beside mean deformation a dominant production mechanism of turbulent kinetic energy is due to the joint effects of non-uniform fluid concentrations and pressure gradients. The (stochastic) density field is an active scalar in the momentum equation.
\item \emph{Turbulence asymmetry:} At high Atwood numbers the mixing process becomes asymmetric. Fluid particles representing different instantaneous density (and thus different inertia) respond very differently to pressure gradients.\cite{Livescu_09b} As a consequence, in a homogeneous Rayleigh-Taylor flow an initially symmetric density distribution quickly develops a sizeable skewness.\cite{Livescu_07,Livescu_08} Similarly, the non-Boussinesq behavior of the pressure leads to asymmetry in the profiles of the mean velocity and the turbulent kinetic energy across inhomogeneous RT mixing layers.\cite{Livescu_09c} In contrast the Boussinesq case remains symmetric.
\item \emph{Non-equilibrium:} The flow evolution is highly non-equilibrium as the production-to-dissipation ratio ranges from hundreds to nearly zero.\cite{Livescu_07}
\item \emph{Transitional:} The Rayleigh-Taylor mixing layer transitions to fully developed turbulence starting from a quiescent state.
\item \emph{Non-stationary:} At no point in time can the flow be deemed statistically stationary.
\item \emph{Anisotropic:} Due to the large external acceleration force, anisotropy is important at both large and small scales.\cite{Livescu_09}
\end{enumerate}
The fundamental physics of the flow evolution must be captured at several different stages. As an example, in an \emph{inhomogeneous Rayleigh-Taylor} flow,\cite{Cabot_06,Livescu_09} two pure fluids are initially separated by a thin perturbed interface in a \emph{quiescent state:} the heavy fluid lies on top of the light one with gravity acting downwards. Due to the unstable configuration, bubbles and spikes grow and penetrate into each other in a \emph{laminar stage,} which then break up into smaller Kelvin-Helmholtz-like eddies. The flow \emph{transitions} into a \emph{fully developed turbulent} mixing layer with its width growing as long as pure fluid is entrained at the edges. Another example is the \emph{homogeneous Rayleigh-Taylor} mixing layer,\cite{Livescu_07,Livescu_08} in which the process starts from a \emph{quiescent state} with random blobs of two pure fluids (in a closed computational box), corresponding to the center of an inhomogeneous RT layer. Potential energy is gradually converted to kinetic energy during \emph{transition} to a \emph{fully developed turbulent} state. After reaching a peak in kinetic energy, a \emph{viscous decay} due to small scale dissipation ensues.

\textbf{Modeling challenges of VD turbulence.} From the viewpoint of statistical turbulence modeling the main challenges are:
\begin{itemize}
\item \emph{To devise a mutually consistent representation for hydrodynamics and mixing.} The model must correctly take into account the effect of the active scalar, the density, on the velocity field, and the mixing of the scalar by the turbulent velocity field.
\item \emph{To represent very different flow-evolutionary stages within the same method.} The model must evolve the flow from a quiescent initial condition through an initially laminar stage, via the highly non-linear process of transition to fully developed turbulence, followed by a possible decay if pure fluids are no longer entrained.
\item \emph{To provide a higher level statistical description than current VD moment closures.} We require the model to predict the full one-point one-time probability distribution function of the active mixing field (the density), due to its fundamental role in mixing-driven VD flows.
\item \emph{To formulate a simple method that captures the essentials.} The numerical method must be computationally inexpensive, so that it can be used as an engineering closure.
\end{itemize}

PDF methods seek to compute the one-point joint PDF of the fluctuating flow variables. Reviews on PDF methods for turbulent flows are provided by Pope,\cite{Pope_85} Kollmann,\cite{Kollmann_90} Dopazo\cite{Dopazo_94} and by Haworth.\cite{Haworth_09} The most widely used velocity PDF models are variants of \Eqre{eq:orGLM}: by specifying $G_{ij}$ and $C_0$ differently several models have been constructed (see e.g.\ Ref.\ \onlinecite{Pope_94}) that are routinely used today in combustion simulations.\cite{Haworth_09} To highlight the challenge of PDF methods for VD turbulence, the fundamental assumptions of the standard PDF methodology for hydrodynamics, \Eqre{eq:orGLM}, are enumerated as follows:
\begin{enumerate}
\item \emph{Uniform-density formulation.} \Eqre{eq:orGLM} has originally been developed for constant-density flows, $\varrho_0=\mathrm{const}$. In low-speed combustion the fluid density is treated as a function of the species concentrations and temperature, $\varrho=\varrho(\bv{y},T)$, and the mean density, $\rmean{\varrho}$, is used in place of $\varrho_0$. This non-constant, but from the viewpoint of the momentum equation still uniform-density case, does not take direct effects of a \emph{stochastic} variable density field on the particle momentum into account. This is most easily seen in the pressure-gradient term of \Eqre{eq:orGLM} and less directly in the other two terms which will be discussed in more detail.
\item \emph{Small scale isotropy.} With the isotropic stochastic diffusion term, $\delta_{ij}\mathrm{d}W_j$, \Eqre{eq:orGLM} adheres to Kolmogorov's hypothesis of local isotropy \emph{by construction.} Situations where the hypothesis is known to fail are non-equilibrium, highly-distorted, moderate-\textit{Re} or VD flows. The fluctuating velocity field in these cases may exhibit significant level of anisotropy at the small scales.\cite{Livescu_09c,Chung_10,Gualtieri_10}
\item \emph{Fully developed turbulence.} \Eqre{eq:orGLM} assumes a fully developed, i.e.\ high-\textit{Re}, flow with isotropic inertial range scaling.\cite{Haworth_86}
\end{enumerate}
The above general discussion makes it clear that developing a PDF model for VD flows is a largely unexplored territory. Due to the wide variety of challenges associated with non-stationary variable-density flows, the development is documented in a series of three papers, as described in the following subsection. This article is the second in the series.

\subsection{Objectives of the article}
Our aim is to develop a model that represents the temporally evolving joint PDF of density and velocity in variable-density pressure-gradient-driven turbulence. The development is carried out in three parts:
\begin{enumerate}
\item \emph{Development of a VD material mixing model.} A previous article,\cite{Bakosi_10} devoted to the density equation, addressed the challenges associated with active scalar mixing in VD flows and developed a Lagrangian stochastic equation to represent conservation of mass.
\item \emph{Development of the Langevin model for the velocity field in VD flows.} The extension of the momentum equation from constant-density shear-driven flows to VD-PGDT is the subject of this paper. Two critical ingredients and their rigorous mathematical consequences will be discussed:
\begin{enumerate}
  \item The representation of the instantaneous density field in the momentum equation, resulting in an appearance of the mass flux (and thus turbulence production) that does not require explicit modeling; and,
  \item The introduction of a new tensorial stochastic diffusion term representing small scale anisotropy.
\end{enumerate}
\item \emph{Joint PDF model specification and validation.} The strongly coupled nature of active material mixing and hydrodynamics in VD-PGDT will require a joint model. This is the subject of a forthcoming article,\cite{Bakosi_10c} which will combine the ideas about mixing and hydrodynamics into a comprehensive modeling framework.
\end{enumerate}

\subsection{Navier-Stokes equation for variable-density flow}
It seems appropriate to specify, at the outset, the dynamic level of hydrodynamical approximation that is to be modeled.

We consider variable-density flows, where differences in fluid density arise solely from non-uniform mass concentrations, e.g.\ due to mixing of different fluids. Density variations originating from pressure and temperature changes are neglected at this time, $\varrho(\bv{y},p,T)=\varrho(\bv{y})$.

For this class of variable-density flows, the Navier-Stokes equation, governing the instantaneous fluid particle velocity, $v_i$, in the presence of viscous and external forces, is written symbolically as
\begin{equation}
\mathrm{d}v_i = \left(g_i - v p,_i + \mu v\nabla^2v_i\right)\mathrm{d}t,\label{eq:vns}
\end{equation}
with the specific volume,
\begin{equation}
v = 1/\varrho.
\end{equation}
Here $g_i$, $p$, $\varrho$ and $\mu$ denote the acceleration force per unit mass, the pressure, the density and the constant dynamic viscosity, respectively. Although in VD flows the viscosity is non-uniform, at this time we take $\mu\!\approx\!\textrm{const.}$ as we are interested in exploring the new simpler physics (VD turbulence) without the complicating issues of non-uniform viscosity. Accordingly, in \Eqre{eq:vns} the viscous force is approximated as $v [\mu(v_{i,j}+v_{j,i}) - 2/3\mu v_{k,k}\delta_{ij}],_j \approx \mu v\nabla^2v_i$.

The constant-density counterpart of \Eqre{eq:vns}, to which the Langevin equation \Eqrs{eq:orGLM} is a model in turbulent flows, reads
\begin{equation}
\mathrm{d}v_i = \left(g_i - p,_i/\varrho_0 + \mu/\varrho_0\nabla^2v_i\right)\mathrm{d}t.\label{eq:ns}
\end{equation}
In the following we will denote a constant density by $\varrho_0$ and the variable one by $\varrho$.

\subsection{Outline of the paper}
To set the stage, the generalized Langevin model, \Eqre{eq:orGLM}, developed for shear-driven constant-den\-sity flows, is reviewed in Sec.\ \ref{sec:GLM}. This is to point out its most important underlying assumptions and so that the later development can highlight the major differences in VD flows. The Langevin equation is extended to VD-PGDT in Sec.\ \ref{sec:vPDF-eqspec}. Sec.\ \ref{sec:vPDF-moments} investigates the new features of the variable-density SDE by deriving its ensemble moment equations and comparing them to those derived from the Navier-Stokes equation. Sec.\ \ref{sec:Favremoments} does the same for the Favre moments. This is to put the proposed PDF formulation into context with Favre VD moment closures. Sec.\ \ref{sec:BoussinesqLimit} discusses the VD PDF model in the Boussinesq limit. The important characteristics of the formulation and the main results are summarized in Sec.\ \ref{sec:vPDF-summary}.

\section{Review of the Langevin model for constant-density shear flows}
\label{sec:GLM}
This section reviews the generalized Langevin model and its underlying assumptions.

Decomposing the velocity and pressure into mean and fluctuating parts, the Navier-Stokes equation for constant-density flows, \Eqre{eq:ns}, takes the form
\begin{align}
\begin{split}
\mathrm{d}v_i & = \left(g_i - \rmean{p},_i/\varrho_0 + \mu/\varrho_0\nabla^2\rmean{v}_i\right)\mathrm{d}t\\
&\quad + \left(-p\rf\!,_i/\varrho_0 + \mu/\varrho_0\nabla^2v_i\rf\right)\mathrm{d}t,\label{eq:lve}
\end{split}
\end{align}
with the fluctuation defined as $y\rf = y-\rmean{y}$.

A widely used model equation for the particle velocity, $v_i^*$, in constant-density shear flows is the \emph{generalized Langevin model} (GLM),\cite{Haworth_86}
\begin{align}
\begin{split}
\mathrm{d}v^*_i & = \left(g_i - \rmean{p},_i/\varrho_0 + \mu/\varrho_0\nabla^2\rmean{v}_i\right)\mathrm{d}t\\
&\quad + G_{ij}\left(v^*_j-\rmean{v}_j\right)\mathrm{d}t + \left(C_0\varepsilon\right)^{1/2}\mathrm{d}W_i.
\label{eq:GLM}
\end{split}
\end{align}
In the following, the star superscript ($^*$) will denote a model for an instantaneous quantity, such as $v^*_i$ for $v_i$.

In \Eqre{eq:GLM} $G_{ij}$ is a second-order tensor which, in shear flows, is assumed to depend on local values of the Reynolds stress, $\rmean{v\rf_iv\rf_j}$, the dissipation rate of turbulent kinetic energy, $\varepsilon$, and the mean velocity gradient, $\rmean{v}_{i,j}$. $C_0$ is a positive constant and $\mathrm{d}W_i(t)$ is a vector-valued Wiener process\cite{Gardiner_09} with zero mean and covariance $\rmean{\mathrm{d}W_i\mathrm{d}W_j}=\mathrm{d}t\delta_{ij}$. The statistics on the right hand side of \Eqre{eq:GLM} are understood to be evaluated at the particle position, $x_i^*$, governed by
\begin{equation}
\mathrm{d}x_i^* = v_i^*\mathrm{d}t.
\label{eq:x}
\end{equation}
Three fundamental assumptions underly \Eqre{eq:GLM}:
\begin{enumerate}
\item \emph{The Markov property},\cite{Gardiner_09} which assumes that the velocity can be described by a Fokker-Planck equation and, equivalently, by a SDE, such as \Eqre{eq:GLM}. Strictly speaking, turbulence is not a Markovian process, but the fluctuations in the inertial subrange, for which \Eqre{eq:GLM} has been developed, can be closely approximated by Markov processes.\cite{Monin_Yaglom_75}
\item \emph{Linear dependence}: through the linear drift term $G_{ij}(v^*_j-\rmean{v}_j)\mathrm{d}t$ the effect of fluctuations of the surrounding fluid is modeled as a linear function of the velocity. This is inconsistent with the quadratic dependence of the fluctuating pressure gradient on the fluctuating velocity.\cite{Kollmann_90} However, linearity (as an approximation) may be justified by (1) the correct behavior of the equation in homogeneous shear turbulence, i.e.\ an arbitrary velocity PDF relaxes to a joint normal, and (2) that realizability of the Reynolds stress tensor is automatically satisfied provided that $C_0$ is non-negative and $G_{ij}$ and $C_0$ are bounded.\cite{Pope_85,Durbin_Speziale_94}
\item \emph{Local isotropy}: the isotropy of the diffusion term $(C_0\varepsilon)^{1/2}\delta_{ij}\mathrm{d}W_j$ implies isotropy at the small scales, thereby ensuring consistency with Kolmogorov's hypothesis of local isotropy.\cite{Obukhov_59,Monin_Yaglom_75}
\end{enumerate}
Since the first drift term of \Eqre{eq:GLM} represents the effects of the mean forces on the particle and the last two terms do not affect the mean, the equation is consistent with the mean of the Navier-Stokes equation \Eqrs{eq:ns}.\cite{Thomson_87,Pope_87} A direct consequence of the linearity of the second drift term in $v_i^*$ \emph{and} the independence of $v_i^*$ of the stochastic term is Gaussianity of the joint velocity PDF.\cite{Gardiner_09}

Comparing \Eqres{eq:lve} and \Eqrs{eq:GLM} shows that the last two terms in \Eqre{eq:GLM} jointly model the combined effect of the fluctuating pressure gradient and viscous dissipation:
\begin{equation}
\begin{split}
& \left(-p\rf\!,_i/\varrho_0 + \mu/\varrho_0\nabla^2v_i\rf\right)\mathrm{d}t =\\
& \qquad\qquad = G_{ij}\left(v^*_j-\rmean{v}_j\right)\mathrm{d}t + \left(C_0\varepsilon\right)^{1/2}\mathrm{d}W_i.
\end{split}
\end{equation}

An important consistency condition on the coefficients $G_{ij}$ and $C_0$ is\cite{Haworth_86}
\begin{equation}
\left(1 + \frac{3}{2}C_0\right)\varepsilon + G_{ij}\rmean{v\rf_iv\rf_j} = 0,
\label{eq:cGC}
\end{equation}
which ensures that no spurious turbulent kinetic energy, $k\!=\!\rmean{v_i\rf v_i\rf}/2$, is created in homogeneous turbulence by the model. \Eqre{eq:cGC} is obtained from comparing the evolution equations that govern $k$ in homogeneous turbulence: the one derived from the Langevin model, \Eqre{eq:GLM}, and the one from the Navier-Stokes equation \Eqrs{eq:ns}. The constraint in \Eqre{eq:cGC} ensures the same form for the $k$ equation:
\begin{equation}
\left.\frac{\partial k}{\partial t}\right|_\mathrm{GLM} = \left.\frac{\partial k}{\partial t}\right|_\mathrm{NS} = \mathcal{P}-\varepsilon,\label{eq:kconsistent}
\end{equation}
where $\mathcal{P}$ is the shear production. The simplest way to define $G_{ij}$ to satisfy \Eqre{eq:cGC} is then
\begin{equation}
G_{ij} = -\left(\frac{1}{2} + \frac{3}{4}C_0\right)\frac{\varepsilon}{k}\delta_{ij},
\label{eq:SLM}
\end{equation}
resulting in the \emph{simplified Langevin model} (SLM),\cite{Haworth_86} which corresponds (at the Reynolds stress level) to Rotta's model of return-to-isotropy.\cite{Rotta_51}

The most important characteristics of the generalized Langevin model have been summarized: (1) the model equation is consistent with the mean Navier-Stokes equation; (2) the mathematical form accommodates Kol\-mo\-go\-rov's hypothesis of local isotropy; (3) a realizable Reynolds stress model is ensured, provided the coefficients satisfy certain mild conditions; and (4) the predicted joint velocity distribution is Gaussian.

\section{A Langevin model for variable-density pressure-gradient-driven flows}
\label{sec:vPDF-eqspec}
This section extends the Langevin model, \Eqre{eq:GLM}, to VD-PGDT and discusses its main ingredients.

We propose to model the Lagrangian velocity increment in VD turbulence by
\begin{align}
\mathrm{d}v^*_i & = \left(g_i - \rmean{p},_i/\varrho^* + \mu/\varrho^*\nabla^2\rmean{v}_i\right)\mathrm{d}t + G_{ij}\left(v^*_j-\fmean{v_j}\right)\mathrm{d}t\nonumber\\
&\quad + \left(\phi_{\scriptscriptstyle I}\varepsilon\right)^{1/2}\mathrm{d}W_i + \left(\phi_{\scriptscriptstyle D}\varepsilon\right)^{1/2}h_{ij}\mathrm{d}W'_j,\label{eq:TLM}
\end{align}
with positive bounded tensorial diffusion $h_{ij}$, bounded functions $\phi_{\scriptscriptstyle I}>0$ and $\phi_{\scriptscriptstyle D}>0$, and independent Wiener processes $\mathrm{d}W_i$ and $\mathrm{d}W'_j$.

The SDE \Eqrs{eq:TLM} differs from \Eqrs{eq:GLM} in three distinct ways, subsequently discussed in more detail in the following three subsections:
\begin{enumerate}
\renewcommand{\theenumi}{\emph{\Alph{enumi}}}
\item \emph{Instantaneous density.} \Eqre{eq:TLM} is intended to be coupled to a stochastic equation governing the instantaneous density field, $\varrho^*$, representing conservation of mass.
\item \emph{Relaxation to Favre-averaged velocity.} The linear relaxation term, $G_{ij}(v^*_j-\fmean{v_j})\mathrm{d}t$, involves the Favre-averaged velocity, $\fmean{v_i} = \rmean{\varrho v_i}/\rmean{\varrho}$.
\item \emph{Small scale anisotropy.} A new tensorial diffusion term, $(\phi_{\scriptscriptstyle D}\varepsilon)^{1/2}h_{ij}\mathrm{d}W'_j$, is introduced. The constant, $C_0$, in the original isotropic diffusion term is exchanged to $\phi_{\scriptscriptstyle I}$ which together with $\phi_{\scriptscriptstyle D}$ will be specified later.
\end{enumerate}

\subsection{Instantaneous density}
In VD turbulence large density variations play a major role and, compared to the Boussinesq case, the density fluctuations can no longer be neglected in the inertia terms of the Navier-Stokes equation. Extending the SDE \Eqrs{eq:GLM} to VD flows requires a representation of the variable density field. \Eqre{eq:TLM} accomplishes this in the most profitable way allowed by one-point PDF methods: a stochastic density equation, such as discussed in Ref.\ \onlinecite{Bakosi_10}, is coupled at the \emph{instantaneous} (particle) level.

Comparing the constant-density and variable-density models, \Eqres{eq:GLM} and \Eqrs{eq:TLM}, we see that in the VD case the particle density, $\varrho^*$, divides the mean pressure gradient and the large-scale viscous terms,
\begin{align}
-\rmean{p},_i/\varrho_0 \quad & \Rightarrow \quad -\rmean{p},_i/\varrho^*,\label{eq:mf}\\
\mu/\varrho_0\nabla^2\rmean{v}_i \quad & \Rightarrow \quad \mu/\varrho^*\nabla^2\rmean{v}_i,\label{eq:lsd}
\end{align}
which has the following consequences:
\begin{itemize}
\item \emph{The full density PDF is coupled to the mean forces in the momentum equation.} Since the governing equation for $\varrho^*$ provides the full density PDF, the mixing state is represented by including the effects of all density moments. As Ref.\ \onlinecite{Bakosi_10c} will demonstrate, this is crucial in capturing the asymmetric PDF of the fluid density at high Atwood numbers. \Eqres{eq:mf} and \Eqrs{eq:lsd} ensure the coupling of the density PDF to (i.e.\ the effects of all its moments on) the mean forces.
\item \emph{The effects of the mass flux on the Reynolds stress appear closed.} As will be shown in Sec.\ \ref{sec:vPDF-moments}, \Eqre{eq:mf} is the key to represent the effect of the mass flux on the Reynolds stress in closed mathematical form. This is crucial in variable-density pressure-gradient-driven flows, as the mass flux relates to a primary mechanism of kinetic energy production, which in moment closures requires additional model equations.
\end{itemize}

The Navier-Stokes equation for VD flows, \Eqre{eq:vns}, with the velocity and pressure decomposed, is
\begin{equation}
\mathrm{d}v_i = \left(g_i - v\rmean{p},_i + \mu v\nabla^2\rmean{v}_i\right)\mathrm{d}t + \left(-vp\rf\!,_i + \mu v\nabla^2v_i\rf\right)\mathrm{d}t,\label{eq:vlve}
\end{equation}
where the instantaneous specific volume, $v=\rmean{v}+v\rf$, is \emph{not decomposed.} Comparing \Eqres{eq:TLM} and \Eqrs{eq:vlve} shows that the combined effects of the fluctuating pressure gradient and viscous dissipation (which now are multiplied by the instantaneous specific volume, $v=1/\varrho$) are jointly modeled as
\begin{equation}
\begin{split}
& \left(-vp\rf\!,_i + \mu v\nabla^2v_i\rf\right)\mathrm{d}t = G_{ij}\left(v^*_j-\fmean{v_j}\right)\mathrm{d}t\\
&\quad\qquad\qquad + \left(\phi_{\scriptscriptstyle I}\varepsilon\right)^{1/2}\mathrm{d}W_i + \left(\phi_{\scriptscriptstyle D}\varepsilon\right)^{1/2}h_{ij}\mathrm{d}W'_j.
\end{split}\label{eq:model}
\end{equation}
In other words, in \Eqre{eq:TLM} the last three terms are a model, while the representation of the mean forces (modulated by the instantaneous specific volume, $v^*=1/\varrho^*$), $g_i - \rmean{p},_i/\varrho^* + \mu/\varrho^*\nabla^2\rmean{v}_i$, is exact. This is a consequence of the availability of the one-point density PDF, represented here via the instantaneous $\varrho^*$. Although the governing equation of $\varrho^*$ may contain modeling,\cite{Bakosi_10} the \emph{coupling} of $\varrho^*$ to the mean forces terms of the momentum SDE \Eqrs{eq:TLM} is mathematically exact. As long as the joint density-velocity PDF is valid (whose definition is discussed later) and the marginal density PDF ensures conservation of mass, the statistics involving the density, such as $\rmean{\varrho\rf v\rf_i}$, are finite, consistent and physically realizable. This is discussed further in Sec.\ \ref{sec:vPDF-moments}. In this paper, we do not assume a particular functional form for $\varrho^*$, only its availability and validity. Ref.\ \onlinecite{Bakosi_10} discusses one choice of the density equation, which can be used in conjunction with \Eqre{eq:TLM}. However, we emphasize that the current development is independent of the functional form of the density model.

\subsection{Relaxation to Favre-averaged velocity}
Comparing the constant-density and variable-density SDEs, \Eqrs{eq:GLM} and \Eqrs{eq:TLM}, shows that the linear relaxation term in the VD case involves the Favre-average of the velocity field, $\fmean{v_i} = \rmean{\varrho v_i}/\rmean{\varrho}$ as
\begin{equation}
G_{ij}\left(v^*_j-\rmean{v}_j\right) \quad \Rightarrow \quad G_{ij}\left(v^*_j-\fmean{v_j}\right).\label{eq:fsv}
\end{equation}
The same approach has been taken by Delarue \& Pope\cite{Delarue_97} to develop a PDF model for high-speed compressible shear flows. In the application of \Eqre{eq:GLM} to turbulent combustion relaxation to the Favre average is a standard procedure. Sec.\ \ref{sec:vPDF-moments} will show that \Eqre{eq:fsv} is instrumental in representing the effects of the fluctuating specific volume on the mean velocity, important in mixing flows with large density differences.

\subsection{Small scale anisotropy}
In high-Reynolds-number shear flows it is generally assumed that the small scales become isotropic and independent of the (anisotropic) large scales where most of the turbulence production takes place. Exceptions are strongly distorted flows\cite{Gualtieri_10} or, as recently shown by Livescu \& Ristorcelli,\cite{Livescu_08} Livescu et al.\cite{Livescu_09,Livescu_09c} and Chung \& Pullin,\cite{Chung_10} the Rayleigh-Taylor mixing layer. In RT flows the buoyancy force has a significant effect on the smallest scales, resulting in small scale anisotropy. This is apparent in the Reynolds stress anisotropy, $b_{ij}(\Hat{\kappa})\!=\!\rmean{v\rf_iv\rf_j}/\rmean{v\rf_kv\rf_k}-\delta_{ij}/3$ at high wavenumbers $\Hat{\kappa}$,\cite{Livescu_09c,Chung_10} and in the dissipation rate anisotropy, $d_{ij}\!=\!\varepsilon_{ij}/\varepsilon_{kk}\!-\!\delta_{ij}/3$, Ref.\ \onlinecite{Livescu_08}. The above studies show that this is prevalent in the Atwood number range $A\!=\!0.04\sim0.75$. We are interested in the full Atwood number range of $0\!<\!A\!<\!1$. Consequently, assuming small scale isotropy, $d_{ij}\approx0$, in such pressure-gradient-driven flows is not justified. Accounting for the dissipation rate anisotropy is crucial in predicting the correct Reynolds stress tensor, whose budget is directly affected by $d_{ij}$ via the small scale dissipation term $\varepsilon_{ij}\!=\!2\varepsilon(d_{ij}+\delta_{ij}/3)$, see also \Eqre{eq:ersd}. The anisotropic behavior of $\varepsilon_{ij}$, governed by $d_{ij}$, is responsible for different dissipation rates of the individual components of the Reynolds stress.

In the constant-density model SDE \Eqrs{eq:GLM}, the small scales are isotropic by construction, which is built into the stochastic term, $(C_0\varepsilon)^{1/2}\delta_{ij}\mathrm{d}W_j$, in accord with Kolmogorov's hypothesis. As will be shown in Sec.\ \ref{sec:vPDF-moments}, in the absence of shear production, a source of (single-point) anisotropy in $b_{ij}$ can be represented by an anisotropic specification for $G_{ij}$, while both $b_{ij}(\Hat{\kappa})\!\approx\!0$ and $d_{ij}(\Hat{\kappa})\!\approx\!0$ at the small scales. As PGDT exhibits anisotropy at both large and small scales (in both $b_{ij}$ and $d_{ij}$), the variable-density velocity model, \Eqre{eq:TLM}, relaxes the third assumption implied by \Eqre{eq:GLM} and abandons the consistency with Kolmogorov's hypothesis of small scale isotropy. It replaces the isotropic diffusion term by the sum of an isotropic and an anisotropic (tensorial) diffusion term as
\begin{equation}
\left(C_0\varepsilon\right)^{1/2}\mathrm{d}W_i \quad \Rightarrow \quad \left(\phi_{\scriptscriptstyle I}\varepsilon\right)^{1/2}\mathrm{d}W_i + \left(\phi_{\scriptscriptstyle D}\varepsilon\right)^{1/2}h_{ij}\mathrm{d}W'_j.\label{eq:diff}
\end{equation}
The first term in \Eqre{eq:diff}, proportional to $\phi_{\scriptscriptstyle I}$, is designed to account for the effects of the isotropic part of the kinetic energy production/dissipation, while the second one, proportional to $\phi_{\scriptscriptstyle D}$ and $h_{ij}$, for the effects of its deviatoric part. As will be shown in Sec.\ \ref{sec:vPDF-moments}, Eqs.\ (\ref{eq:mrs}--\ref{eq:incmrs}), the stochastic diffusion terms in \Eqre{eq:diff} are source/sink terms, depending on the sign of the given Reynolds stress component.

The tensorial term in \Eqre {eq:diff} is assumed to be proportional to $\varepsilon\!=\!\varepsilon_{kk}/2$. This is a conventional procedure in low-Reynolds-number flows, such as near walls,\cite{Rotta_51,Launder_83} and justified here by $d_{ij}(t)$ being non-zero through all the Rayleigh-Taylor flow evolution, see Fig.\ 16 in Ref.\ \onlinecite{Livescu_08}. In other words, the scalar kinetic energy dissipation rate, $\varepsilon$, is made anisotropic by the tensor $h_{ij}$ in the VD model. This ensures that the small scales are anisotropic: the Lagrangian velocity structure function of the process governed by \Eqre{eq:TLM} becomes\cite{Pope_00}
\begin{equation}
\begin{split}
D_{ij}(s) & \equiv \rmean{[v_i^*(t+s)-v_i^*(t)][v_j^*(t+s)-v_j^*(t)]}\\
& = (\phi_{\scriptscriptstyle I}\delta_{ij} + \phi_{\scriptscriptstyle D}h_{ik}h_{kj})\varepsilon s,
\end{split}
\end{equation}
with $s\!\ll\!\tau\!\ll\!T$, where $\tau$ and $T\!\equiv\!\left\|\rmean{v}_{k,k}\right\|^{-1}$ denote the time scales of the dissipation and the mean deformation, respectively. As will be shown in Sec.\ \ref{sec:vPDF-moments}, $h_{ij}$, as a contribution of the small scales, provides an additional source (beside $G_{ij}$) in the budget of the one-point Reynolds stress anisotropy, $b_{ij}$.

\textbf{Constraints.} Up to this point, the functional forms of the diffusion terms for VD flows, given by \Eqre{eq:diff}, are formulated based on physical insight and mathematical consistency. These are:
\begin{itemize}
\item \emph{Small scale anisotropy.} The anisotropy of RT flows, indicated by both $b_{ij}\!\ne\!0$ and $d_{ij}\!\ne\!0$ at the small scales.
\item \emph{Correct unit.} As $\phi_{\scriptscriptstyle I}$ and $\phi_{\scriptscriptstyle D}^{1/2}h_{ij}$ are assumed to be non-dimensional, each component of the diffusion terms has the unit as that of $\sqrt{\varepsilon\mathrm{d}t}$.
\item \emph{Second-order tensor coefficient.} The simplest way to introduce directional dependence into the joint statistics of the fluctuating velocity components is to do so via the product of a second-order tensor and a vector-valued Wiener process.
\end{itemize}
The coefficients $\phi_{\scriptscriptstyle I}$, $\phi_{\scriptscriptstyle D}$ and $h_{ij}$ must be specified based on the following considerations:
\begin{enumerate}
\item \emph{Consistency with the SDE.} The coefficient tensor should be a square root of a symmetric and non-negative semi-definite tensor, $H_{ij}\!=\!h_{ik}h_{jk}$. This is required mathematically for a SDE, such as \Eqre{eq:TLM}, to represent a diffusion.\cite{vanKampen_04}
\item \emph{Reflect the physical source of small scale anisotropy.} The diffusion coefficient tensor, $h_{ij}$, should be a function of the given source of small scale anisotropy. In buoyantly driven flows this may be the buoyancy force, $g_i$, in other types of externally accelerated or strongly distorted flows, the body force, the pressure-gradient force or the mean strain rates, responsible for the anisotropy. This ensures that the model directly represents the effect of the source of anisotropy on the small scales.
\item \emph{Correct distribution.} None of $h_{ij}$, $\phi_{\scriptscriptstyle I}$ or $\phi_{\scriptscriptstyle D}$ should be an explicit function of the particle velocity, $v_i^*$, if the Gaussianity of the joint velocity PDF is to be preserved. In constant-density homogeneous shear flows the velocity distribution is very close to a joint normal.\cite{Tavoularis_81} Low-Atwood-number direct numerical simulations (DNS) show that joint Gaussianity of the velocity components is also a good approximation in RT flows.\cite{Vladimirova_09} As the drift of the Langevin equation \Eqrs{eq:GLM} is linear and its diffusion is not an explicit function of $v_i^*$, it is an Ornstein-Uhlenbeck process,\cite{Gardiner_09} whose statistically stationary solution is a Gaussian. As an asymptotic requirement of the VD model \Eqre{eq:TLM}, that is to reduce to the constant-density Langevin model, the new diffusion terms should approach this attribute as a limit when $\varrho\rf\to0$.
\item \emph{Asymptotic small scale isotropy.} As an extension of the model for shear flows, the new model is required to reduce to small scale isotropy if its source vanishes.
\item \emph{Correct one-point anisotropy.} The combined effects of the new stochastic diffusion terms with $\phi_{\scriptscriptstyle I}$, $\phi_{\scriptscriptstyle D}$, $h_{ij}$ and $G_{ij}$ should be the correct level of one-point anisotropy in the correct components of $b_{ij}$.
\item \emph{Consistent turbulent kinetic energy.} Introducing anisotropy at the small scales must not create spurious turbulent kinetic energy. In other words, the kinetic energy budgets of the VD PDF model and of the Navier-Stokes equation must be consistent with each other, similarly to \Eqre{eq:kconsistent} in constant-density flows.
\end{enumerate}

\subsection{Summary}
The extension of the constant-density Langevin model to variable-density pressure-gradient-driven flows comprises of (1) working with the instantaneous mass-density, (2) relaxation towards the Favre mean (instead of the Reynolds mean) velocity, and (3) accounting for small scale anisotropy with a new tensorial diffusion term (instead of built-in local isotropy).

In the next sections we examine the equations governing the moments of the VD SDE \Eqrs{eq:TLM}. The derived moment equations are rigorous mathematical consequences of the particular functional form of \Eqre{eq:TLM}, which establish constraints on the coefficients, $G_{ij}$, $h_{ij}$, $\phi_{\scriptscriptstyle I}$ and $\phi_{\scriptscriptstyle D}$, that serve as a guideline for model specification.

\section{Reynolds moment equations}
\label{sec:vPDF-moments}
To investigate the new features of the SDE \Eqrs{eq:TLM}, and to aid the specification of its coefficients, the evolution equations for the first two moments of velocity, derived from the PDF model and the VD Navier-Stokes equation, are compared and examined. Here we discuss \emph{ensemble} (or Reynolds) averaged statistics as they explicitly show the effects of the density fluctuations. Sec.\ \ref{sec:Favremoments} discusses the Favre moments.

\subsection{The PDF transport equation}
This section gives the transport equation from which the equations governing the Reynolds moments are derived.

The equivalent Fokker-Planck equation governing the Eulerian joint PDF of density and velocity, $f(\varrho,\bv{v};\bv{x},t)$, is derived from the VD Langevin model, \Eqre{eq:TLM} and \Eqre{eq:x}, see Ref.\ \onlinecite{Pope_85},
\begin{equation}
\begin{split}
\frac{\partial f}{\partial t} + v_i\frac{\partial f}{\partial x_i} & = - \frac{\partial}{\partial v_i}\Big[\left(g_i - \rmean{p},_i/\varrho + \mu/\varrho\nabla^2\rmean{v}_i\right) f\Big]\\
&\quad - \frac{\partial}{\partial v_i}\Big[G_{ij}\big(v_j - \fmean{v_j}\big) f\Big]\\
&\quad + \frac{1}{2}\frac{\partial^2}{\partial v_i\partial v_j}\Big[\big(\phi_{\scriptscriptstyle I}\delta_{ij}+\phi_{\scriptscriptstyle D}H_{ij}\big)\varepsilon f\Big]\\
&\quad + \textrm{density model terms,}\label{eq:vFP}
\end{split}
\end{equation}
where $H_{ij}$ is a symmetric non-negative semi-definite tensor with its Cholesky-decomposition
\begin{equation}
H_{ij} = h_{ik}h_{kj}.
\end{equation}
Beside specifying the coefficients, $G_{ij}$, $H_{ij}$, $\phi_{\scriptscriptstyle I}$ and $\phi_{\scriptscriptstyle D}$, \Eqre{eq:vFP} can be closed if a model for the dissipation rate, $\varepsilon$, is specified. Such a model is given by van Slooten et al.\cite{vanSlooten_98}

The particular form of the density equation, that yields additional fluxes in the density sample space in \Eqre{eq:vFP}, is not important for our purposes. A density model, based on the beta PDF, is given in Ref.\ \onlinecite{Bakosi_10}. However, the beta model is only one of many possibilities and thus we do not detail the functional form of the density process, $\varrho^*(t)$. \Eqre{eq:vFP} only signals the availability of the instantaneous density field (as a sample space variable, $\varrho$) and thus the full density PDF.

In PDF methods, terms originating from the physical process of advection appear in closed form, thus we will incorporate these in the Lagrangian derivative. As a consequence, \emph{the derived moment equations will represent the rate of change along instantaneous Lagrangian paths,} as specified by
\begin{equation}
\ld{} \equiv \frac{\partial}{\partial t} + v_k\frac{\partial}{\partial x_k}.\label{eq:ld}
\end{equation}
This allows us to concentrate on the fundamental physics due to pressure-gradient and viscous forces and their modeling, separately from advection. It is worth emphasizing that the terms underlying $\ild{(\cdot)}$ may be different in equations governing different statistics.

In a joint PDF formulation for density and velocity, both Favre and Reynolds-averaged statistics can be obtained from the PDF. Variable-density flows are traditionally investigated in the Favre-averaged framework, in which the moment equations take a simpler form at the price of concealing some effects of the density fluctuations. We work here with Reynolds-averaged statistics for three reasons:
\begin{enumerate}
  \item With Reynolds averaging the effects of density fluctuations are made clear.
  \item All new features of the extended model can be investigated through equations for the Reynolds moments, as will be seen.
  \item In the PDF framework modeling is not performed explicitly on the moment equations, nor are these explicitly discretized and solved. Therefore, from the viewpoint of analyzing the features of the velocity model, it is unimportant which type of statistics we investigate.
\end{enumerate}
In the following, we concentrate on the new features of the extended model, \Eqre{eq:TLM}, in VD flows where the density fluctuations are large, compared to the GLM, \Eqre{eq:GLM}, in constant-density flows. The Favre-averaged moments and the Boussinesq limit are discussed in the subsequent sections.

\subsection{The ensemble mean velocity equation: $\rmean{v}_i$}
The modeled Reynolds-averaged mean velocity equation in VD flows is derived from \Eqre{eq:vFP}, see Ref.\ \onlinecite{Pope_85}:
\begin{equation}
\left.\ld{\rmean{v}_i}\right|_\mathrm{SDE} = g_i - \rmean{v}\!\cdot\!\rmean{p},_i + \mu\rmean{v}\!\cdot\!\nabla^2\rmean{v}_i + G_{ij}\rmean{v_j\ff},\label{eq:mmv}
\end{equation}
with the Favre fluctuation of the velocity, $v\ff_i = v_i - \fmean{v_i}$.

For comparison to the constant-density case, the corresponding mean velocity equation, derived from the generalized Langevin model, \Eqre{eq:GLM}, reads
\begin{equation}
\left.\ld{\rmean{v}_i}\right|_\mathrm{GLM} = g_i - \rmean{p},_i/\varrho_0 + \mu/\varrho_0\nabla^2\rmean{v}_i.\label{eq:incmmv}
\end{equation}

Comparing \Eqre{eq:mmv} to its counterpart, derived from the Navier-Stokes equation for VD flows, \Eqre{eq:vns},
\begin{equation}
\left.\ld{\rmean{v}_i}\right|_\mathrm{NS} = g_i - \rmean{v}\!\cdot\!\rmean{p},_i + \mu\rmean{v}\!\cdot\!\nabla^2\rmean{v}_i - \rmean{v\rf p\rf\!,_i} + \mu\rmean{v\rf\nabla^2v_i\rf},\label{eq:emv}
\end{equation}
indicates that the combined effects of the fluctuating specific volume (last two terms) are represented by the variable-density model as
\begin{equation}
-\rmean{v\rf p\rf\!,_i} + \mu\rmean{v\rf\nabla^2v_i\rf} = G_{ij}\rmean{v_j\ff}.\label{eq:mm}
\end{equation}
The comparison of the right hand sides of \Eqres{eq:mmv} and \Eqrs{eq:emv} also shows that employing $\rmean{v}_j$ instead of $\fmean{v_j}$ in \Eqre{eq:TLM} would diminish the effect of $v\rf$ on the mean velocity, since it would imply $-\rmean{v\rf p\rf\!,_i} + \mu\rmean{v\rf\nabla^2v_i\rf} = 0$. This is not justified in VD flows with large density fluctuations. In contrast, \Eqre{eq:mm} indicates how the effects of the fluctuating specific volume are incorporated into the velocity model for VD flows.

To summarize, the product of the tensor $G_{ij}$ and the mass flux, $\rmean{v\ff_i}$, jointly represents the correlations of the specific volume with the pressure gradient and viscous diffusion, \Eqre{eq:mm}. Compared to the constant-density case, this is an additional task, required of $G_{ij}$, in VD flows with large density fluctuations.

\subsection{The ensemble Reynolds stress equation: $\rmean{v_i\rf v_j\rf}$}
The model equation for the ensemble Reynolds stress, $\rmean{v\rf_iv\rf_j}$, in VD flows, derived from \Eqre{eq:vFP}, becomes
\begin{align}
\begin{split}
\left.\ld{\rmean{v_i\rf v_j\rf}}\right|_\mathrm{SDE} & = \mathcal{M}_{ij} + G_{ik}\rmean{v_j\rf v_k\rf} + G_{jk}\rmean{v_i\rf v_k\rf}\\
& \quad + \big(\phi_{\scriptscriptstyle I}\delta_{ij} + \phi_{\scriptscriptstyle D}H_{ij}\big)\varepsilon,\label{eq:mrs}
\end{split}
\end{align}
where the identity $\rmean{v_i\rf v_j\ff}\equiv\rmean{v_i\rf v_j\rf}$ has been used in the terms involving $G_{ij}$. For comparison to the constant-density case, the GLM equation \Eqrs{eq:GLM} yields
\begin{align}
\left.\ld{\rmean{v_i\rf v_j\rf}}\right|_\mathrm{GLM} & = G_{ik}\rmean{v_j\rf v_k\rf} + G_{jk}\rmean{v_i\rf v_k\rf} + C_0\varepsilon\delta_{ij}.\label{eq:incmrs}
\end{align}

In \Eqre{eq:mrs} the tensor $\mathcal{M}_{ij}$ is defined as the integral over the whole sample space of the joint density-velocity PDF,\cite{Pope_85}
\begin{equation}
\begin{split}
\mathcal{M}_{ij} & = -\iint v\rf_kv\rf_l \frac{\partial}{\partial v_i}\Big[\left(-v\rmean{p},_i + \mu v\nabla^2\rmean{v}_i\right)f\Big]\mathrm{d}\varrho\mathrm{d}\bv{v} =\\
&= -\rmean{v\rf v_i\rf}\!\cdot\!\rmean{p},_j - \rmean{v\rf v_j\rf}\!\cdot\!\rmean{p},_i + \mu\big(\rmean{v\rf v_i\rf}\!\cdot\!\nabla^2\rmean{v}_j + \rmean{v\rf v_j\rf}\!\cdot\!\nabla^2\rmean{v}_i\big).\label{eq:M}
\end{split}
\end{equation}
$\mathcal{M}_{ij}$ collects the effects of the specific volume flux, $\rmean{v\rf v\rf_i}$, due to the mean pressure gradient and the mean viscous forces. The products $-\rmean{v\rf v_i\rf}\cdot\rmean{p},_j - \rmean{v\rf v_j\rf}\cdot\rmean{p},_i$ are an important source of turbulence in VD flows and reflect the fact that Lagrangian particles of different-density fluids accelerate very differently in response to pressure gradients.

Since $\varrho v = 1$, the specific volume flux, $\rmean{v\rf v\rf_i}$, appearing in the Reynolds-averaged formulation, is related to the mass flux by
\begin{equation}
\rmean{\varrho}\!\cdot\!\rmean{v\rf v\rf_i} + \rmean{v}\!\cdot\!\rmean{\varrho\rf v\rf_i} + \rmean{\varrho\rf v\rf v\rf_i} = 0.\label{eq:rvi-vvi}
\end{equation}
Consequently, $\rmean{v\rf v\rf_i}$ can be expressed in terms of $\rmean{\varrho\rf v\rf_i}$, and $\mathcal{M}_{ij}$ can be written as
\begin{equation}
\begin{split}
\mathcal{M}_{ij} & = \rmean{v}\left(a_i\rmean{p},_j + a_j\rmean{p},_i\!\right) - \mu\rmean{v}\big(a_i\nabla^2\rmean{v}_j + a_j\nabla^2\rmean{v}_i\big)\\
&\quad + \big(\rmean{\varrho\rf v\rf v\rf_i}\!\cdot\!\rmean{p},_j + \rmean{\varrho\rf v\rf v\rf_j}\!\cdot\!\rmean{p},_i\!\big)/\rmean{\varrho}\\
&\quad - \mu\big(\rmean{\varrho\rf v\rf v\rf_i}\!\cdot\!\nabla^2\rmean{v}_j + \rmean{\varrho\rf v\rf v\rf_j}\!\cdot\!\nabla^2\rmean{v}_i\big)/\rmean{\varrho},
\end{split}\label{eq:Mrho}
\end{equation}
where $a_i=\rmean{\varrho\rf v\rf_i}/\rmean{\varrho}$. All one-point statistics of $\varrho$, $v$ and $v_i$ can be extracted from the joint PDF, $f$, therefore the terms in \Eqre{eq:Mrho} require no explicit closure assumptions in the PDF formulation. The representation of $\mathcal{M}_{ij}$ is consistent, provided the joint density-velocity PDF is valid. The necessary and sufficient conditions that establish the validity of the PDF, $f(\varrho,\bv{v};\bv{x},t)$, are:\cite{Pope_85}
  \begin{enumerate}
  \item \emph{Realizability:}
  \begin{equation}
   f(\varrho,\bv{v};\bv{x},t) \ge 0.
  \end{equation}
  \item \emph{Normalization property:}
  \begin{equation}
  \iint f(\varrho,\bv{v};\bv{x},t) \mathrm{d}\bv{v}\mathrm{d}\varrho = 1.
  \end{equation}
  \item \emph{Conservation of mean mass:} $f(\varrho,\bv{v};\bv{x},t)$ satisfies conservation of mass in the mean,
  \begin{equation}
  \ld{\rmean{\varrho}} = -\rmean{\varrho}\!\cdot\!\rmean{d} - \rmean{\varrho\rf\ivdrf},
  \end{equation}
  with $d=v_{i,i}$.
  \end{enumerate}
In other words, if a mass-density model, satisfying the above three conditions, is coupled to the VD SDE $\Eqrs{eq:TLM}$, the representation of $\mathcal{M}_{ij}$ remains consistent. Such a density model is discussed in Ref.\ \onlinecite{Bakosi_10}.

\Eqre{eq:M} indicates that $\mathcal{M}_{ij}$ vanishes in constant-density flows, as $v\rf=0$. Consistently, $\mathcal{M}_{ij}=0$ in the constant-density model, \Eqre{eq:incmrs}. Its appearance in the VD Reynolds stress budget, \Eqre{eq:mrs}, is the consequence of employing the instantaneous particle density, $\varrho^*$, in the mean forces terms of the SDE \Eqrs{eq:TLM}.

The Favre Reynolds stress equation, derived from the SDE \Eqrs{eq:TLM} and the Navier-Stokes equation \Eqrs{eq:vns}, are discussed in Sec.\ \ref{sec:Favremoments}. As will be shown, in that framework the effects of the mass flux in the first line of \Eqre{eq:Mrho} appear explicitly, while the terms involving triple correlations are embedded in the Favre Reynolds stress, $\fmean{v\ff_iv\ff_j}$.

The equation governing the ensemble Reynolds stress in VD flows, derived from the Navier-Stokes equation \Eqrs{eq:vns}, is
\begin{equation}
\left.\ld{\rmean{v_i\rf v_j\rf}}\right|_\mathrm{NS} = \mathcal{M}_{ij} + \mathcal{R}_{ij} - \mathcal{T}_{ij} - \varepsilon_{ij} + \mathcal{V}_{ij},\label{eq:ersd}
\end{equation}
with the redistributive, transport, dissipative terms, and the triple correlations, respectively,
{\allowdisplaybreaks
\begin{align}
\mathcal{R}_{ij} & = \rmean{v}\!\cdot\!\rmean{p\rf(v\rf\!_{i,j} + v\rf\!_{j,i})},\label{eq:R}\\
\mathcal{T}_{ij} & = \rmean{v}\left[(\rmean{v_i\rf p\rf}),_j + (\rmean{v_j\rf p\rf}),_i - \mu\nabla^2\rmean{v_i\rf v_j\rf}\right],\label{eq:T}\\
\varepsilon_{ij} & = 2\mu\rmean{v}\!\cdot\!\rmean{v\rf\!_{i,k} v\rf\!_{j,k}},\label{eq:E}\\
\mathcal{V}_{ij} & = -\rmean{v\rf(v_i\rf p\rf\!,_j + v_j\rf p\rf\!,_i)} + \mu\rmean{v\rf(v_i\rf\nabla^2v_j\rf + v_j\rf\nabla^2v_i\rf)}.\label{eq:V}
\end{align}}%
Note that production due to mean deformation, $\mathcal{P}_{ij} = \rmean{v_i\rf v_k\rf}\cdot\rmean{v}_{j,k} + \rmean{v_j\rf v_k\rf}\cdot\rmean{v}_{i,k}$, and turbulent transport, $(\rmean{v_i\rf v_j\rf v_k\rf}),_k$, remain hidden in the above development. These terms are in the Lagrangian derivative, $\mathrm{d}\rmean{v_i\rf v_j\rf}/\mathrm{d}t$.

Comparing the model \Eqrs{eq:mrs} and Navier-Stokes \Eqrs{eq:ersd} Reynolds stress equations in VD flows shows that the terms involving $G_{ij}$ and $\varepsilon$ provide a joint model as
\begin{equation}
\begin{split}
&\mathcal{R}_{ij} - \mathcal{T}_{ij} - \varepsilon_{ij} + \mathcal{V}_{ij} =\\
&\qquad = G_{ik}\rmean{v_j\rf v_k\rf} + G_{jk}\rmean{v_i\rf v_k\rf} + \big(\phi_{\scriptscriptstyle I}\delta_{ij} + \phi_{\scriptscriptstyle D}H_{ij}\big)\varepsilon.
\end{split}\label{eq:rsmodel}
\end{equation}
\Eqre{eq:rsmodel} indicates that the terms in $G_{ij}$ and $\varepsilon$ in the SDE \Eqrs{eq:TLM} represent the combined effects of the pressure-strain correlation, turbulent transport, dissipation and the triple correlations. It also shows that $\mathcal{M}_{ij}$ does not require modeling. The model terms of \Eqre{eq:TLM}, the ones involving $G_{ij}$ and $\varepsilon$, do not account for $\mathcal{M}_{ij}$, which appears naturally in the moment equation \Eqrs{eq:mrs}. $\mathcal{M}_{ij}$ is the most important term that drives variable-density pressure-gradient-driven turbulence.

In second-order moment closures for VD flows equations for the first two Favre moments, $\fmean{v_i}$ and $\fmean{v_i\ff v_j\ff}$, are solved. Since only Favre-averages are involved, it is necessary to separately model the normalized mass flux, $a_i\!=\!-\rmean{v_i\ff}$, to compute its effect on the Reynolds stress. One such model is by Besnard et al.,\cite{Besnard_92} which carries two model equations in addition to the Favre mean and Reynolds stress: one for $a_i$ and another one for the density-specific-volume covariance, $\Hat{b}\!=\!-\rmean{\varrho\rf v\rf}$, appearing in the equation for the mass flux, each with its burden of closure assumptions. As \Eqres{eq:M} and \Eqrs{eq:Mrho} show, in the PDF formulation no modeling of $a_i$ and $\Hat{b}$ is required. If the joint PDF is valid, the representation of both $a_i$ and $\Hat{b}$ are consistent, see also Ref.\ \onlinecite{Bakosi_10}. In the PDF framework modeling is performed on the joint PDF (instead of its moments), e.g.\ on the stochastic equations representing $\varrho^*$ and $v_i^*$. In the case of the SDE \Eqrs{eq:TLM} only the effects of the fluctuating pressure gradient and viscous dissipation are modeled, while the effects of the mean forces, from which $\mathcal{M}_{ij}$ originates, are exact, see also \Eqres{eq:model} and \Eqrs{eq:rsmodel}.

To summarize, in contrast to moment closures, the important source of turbulence in VD flows, $\mathcal{M}_{ij}$ \Eqre{eq:Mrho}, is closed in the PDF formulation. Redistribution, turbulent transport, dissipation and the triple correlations of the specific volume and the gradients of velocity and pressure are jointly modeled, \Eqre{eq:rsmodel}.

\subsection{The ensemble turbulent kinetic energy equation: $k$}
Taking one-half the trace of the Reynolds stress equation gives an equation for the turbulent kinetic energy. Contracting \Eqre{eq:mrs} results in the model equation for the Reynolds-averaged $k=\rmean{v\rf_iv\rf_i}/2$ in VD flows as
\begin{equation}
\left.\ld{k}\right|_\mathrm{SDE} = \mathcal{M} + G_{ij}\rmean{v\rf_iv\rf_j} + \left(\frac{3}{2}\phi_{\scriptscriptstyle I} + \phi_{\scriptscriptstyle D}H\right)\varepsilon,
\label{eq:km}
\end{equation}
with $\mathcal{M}=\mathcal{M}_{ii}/2$ and $H=H_{ii}/2$. For comparison to constant-density flows, the GLM \Eqre{eq:GLM} produces the $k$ equation
\begin{equation}
\left.\ld{k}\right|_\mathrm{GLM} = G_{ij}\rmean{v\rf_iv\rf_j} + \frac{3}{2}C_0\varepsilon.
\label{eq:inckm}
\end{equation}

From \Eqre{eq:ersd}, the equation governing $k$ based on the Navier-Stokes equation \Eqrs{eq:vns} is
\begin{equation}
\left.\ld{k}\right|_\mathrm{NS} = \mathcal{M} + \mathcal{R} - \mathcal{T} - \varepsilon + \mathcal{V},\label{eq:ke}
\end{equation}
with the half of the traces of Eqs.\ (\ref{eq:R}--\ref{eq:V}),
{\allowdisplaybreaks
\begin{align}
\mathcal{R} & = \rmean{v}\!\cdot\!\rmean{p\rf\ivdrf},\\
\mathcal{T} & = \rmean{v}\left[(\rmean{v_i\rf p\rf}),_i - \frac{\mu}{2}\nabla^2\rmean{v\rf_iv\rf_i}\right],\\
\varepsilon & = \mu\rmean{v}\!\cdot\!\rmean{v\rf\!_{i,k} v\rf\!_{i,k}},\\
\mathcal{V} & = -\rmean{v\rf v_i\rf p\rf\!,_i} + \mu\rmean{v\rf v_i\rf \nabla^2v_i\rf}.
\end{align}}%
Comparing \Eqres{eq:km} and \Eqrs{eq:ke} shows that both yield the same evolution of $k$ if
\begin{equation}
G_{ij}\rmean{v\rf_iv\rf_j} + \left(1 + \frac{3}{2}\phi_{\scriptscriptstyle I} + \phi_{\scriptscriptstyle D}H\right)\varepsilon = \mathcal{R} - \mathcal{T} + \mathcal{V}.\label{eq:cGCH}
\end{equation}
\Eqre{eq:cGCH} can also be obtained from contracting \Eqre{eq:rsmodel}.

As $\mathcal{M}_{ij}$ appears closed, its trace, $\mathcal{M}$, requires no explicit modeling. \Eqre{eq:cGCH} in inhomogeneous VD flows with small scale anisotropy can be contrasted to \Eqre{eq:cGC} in homogeneous constant-density flows where the small scales are assumed to be isotropic. The differences are: the trace of the anisotropic diffusion coefficient tensor, $H$, pres\-sure-dilatation, $\mathcal{R}$, pressure and viscous transport, $\mathcal{T}$, and the trace of the triple correlations, $\mathcal{V}$.

In constant-density homogeneous flows $\mathcal{R}\!=\!\mathcal{T}\!=\!\mathcal{V}\!=\!0$ and \Eqre{eq:cGC} is a \emph{consistency condition} that ensures that the GLM \Eqre{eq:GLM} creates no spurious turbulent kinetic energy in homogeneous turbulence. In contrast, \Eqre{eq:cGCH} is a \emph{model} for variable-density inhomogeneous flows with small scale anisotropy. As the joint PDF, $f(\varrho,\bv{v})$, contains no information on the fluctuating pressure, its gradient or the velocity derivatives, their correlations require modeling. \Eqre{eq:cGCH} indicates how the specifications of $G_{ij}$, $H_{ij}$, $\phi_{\scriptscriptstyle I}$ and $\phi_{\scriptscriptstyle D}$ yield a combined model for the effects of the pressure-dilatation covariance, the trace of transport and the triple correlations, $\mathcal{R}\!-\!\mathcal{T}\!+\!\mathcal{V}$. $\mathcal{R}$ is important at high turbulent Mach numbers, while $\mathcal{V}$ at high Atwood numbers, see also Sec.\ \ref{sec:BoussinesqLimit}.

To summarize, \Eqre{eq:cGCH} provides a model constraint in inhomogeneous variable-density flows with small scale anisotropy for the effects of transport and scalar effects of variable density.

\subsection{The ensemble Reynolds stress anisotropy equation: $b_{ij}$}
The normalized one-point Reynolds stress an\-i\-so\-tropy
\begin{equation}
b_{ij} = \frac{\rmean{v\rf_iv\rf_j}}{2k} - \frac{1}{3}\delta_{ij},
\end{equation}
is indicative of the fraction of the turbulent kinetic energy in a given component of the Reynolds stress tensor: $-1/3$ indicates no energy, while $2/3$ indicates 100\% energy in the component. Its importance in both constant-density and VD flows can be highlighted by writing the production term of $\rmean{v\rf_iv\rf_j}$ due to mean deformation as
\begin{equation}
\mathcal{P}_{ij} = 2k\left(b_{ik}+\frac{1}{3}\delta_{ik}\right)\rmean{v}_{j,k} + 2k\left(b_{jk}+\frac{1}{3}\delta_{jk}\right)\rmean{v}_{i,k}.\label{eq:P}
\end{equation}
In our formulation, $\mathcal{P}_{ij}$ is hidden in the Lagrangian derivative of the Reynolds stress balance \Eqre{eq:ersd}, in as much as
\begin{equation}
\left.\ld{\rmean{v_i\rf v_j\rf}}\right|_\mathrm{NS} = \frac{\partial\rmean{v\rf_iv\rf_j}}{\partial t} + \rmean{v}_k\frac{\partial\rmean{v\rf_iv\rf_j}}{\partial x_k} + \frac{\partial\rmean{v\rf_iv\rf_jv\rf_k}}{\partial x_k} + \mathcal{P}_{ij}.\label{eq:ldrs}
\end{equation}
\Eqres{eq:P} and \Eqrs{eq:ldrs} show that the correct prediction of both i\-so\-trop\-ic and de\-vi\-ator\-ic parts of $\rmean{v\rf_iv\rf_j}$ are important in representing the correct turbulence levels. To investigate how anisotropy is created and dissipated by the stochastic VD model, the evolution equation governing $b_{ij}$ is investigated in the following.

\noindent \textbf{The general case.} The $b_{ij}$ equation of the VD PDF model is derived based on \Eqres{eq:mrs} and \Eqrs{eq:km}:
\begin{align}
\!\!\!\left.2k\ld{b_{ij}}\right|_\mathrm{SDE} & = \mathcal{M}^\textrm{d}_{ij} + 2k\left[G_{ik}b_{jk} + G_{jk}b_{ik} + \frac{1}{3}(G_{ij}+G_{ji})\right]\nonumber\\
& \quad - \left[2\big(\mathcal{M}+G_{kl}\rmean{v\rf_kv\rf_l}\big) + \big(3\phi_{\scriptscriptstyle I}+2\phi_{\scriptscriptstyle D}H\big)\varepsilon\right] b_{ij}\nonumber\\
& \quad - \frac{2}{3}G_{kl}\rmean{v\rf_kv\rf_l}\delta_{ij} + \phi_{\scriptscriptstyle D}\varepsilon H^\textrm{d}_{ij},
\label{eq:bvd}
\end{align}
with the deviatoric parts of $\mathcal{M}_{ij}$ and $H_{ij}$ denoted by the superscript ``d'':
\begin{align}
\mathcal{M}^\textrm{d}_{ij} & = \mathcal{M}_{ij} - \frac{2}{3}\mathcal{M}\delta_{ij},\\
H^\textrm{d}_{ij} & = H_{ij} - \frac{2}{3}H\delta_{ij}.
\end{align}
\Eqre{eq:bvd} is the most general case of $b_{ij}$ that can be represented by the VD SDE \Eqrs{eq:TLM}: it assumes variable density $\mathcal{M}_{ij}\ne0$, anisotropic $G_{ij}$ and anisotropic small scales $H_{ij}\ne0$. Depending on the approximations that can be made, several simplified forms of \Eqre{eq:bvd} can be obtained.

\noindent\textbf{(1) GLM in constant-density flows.} As a special case, the GLM \Eqre{eq:GLM} for constant-density flows yields
\begin{align}
2k&\left.\ld{b_{ij}}\right|^{\varrho_0}_\mathrm{GLM} = 2k\left[G_{ik}b_{jk} + G_{jk}b_{ik} + \frac{1}{3}(G_{ij}+G_{ji})\right]\nonumber\\
&\qquad - \left(2G_{kl}\rmean{v\rf_kv\rf_l} + 3C_0\varepsilon\right) b_{ij} - \frac{2}{3}G_{kl}\rmean{v\rf_kv\rf_l}\delta_{ij}.
\label{eq:bglm}
\end{align}
This amounts to applying the GLM in constant-density flows, where $G_{ij}$ is given by an an\-i\-so\-trop\-ic model, as discussed in Pope.\cite{Pope_94} It can be obtained from the general case by assuming constant density $\mathcal{M}_{ij}=0$, anisotropic $G_{ij}$ and isotropic small scales $H_{ij}=0$.

\noindent\textbf{(2) SLM in constant-density flows.} \Eqre{eq:bglm} further simplifies if $G_{ij}$ is chosen to be isotropic, e.g.\ in case of the SLM, \Eqre{eq:SLM}, since all terms involving $G_{ij}$ vanish, as expected:
\begin{equation}
\left.\ld{b_{ij}}\right|^{\varrho_0}_\mathrm{SLM} = -\frac{3}{2}C_0\frac{\varepsilon}{k}b_{ij},\label{eq:br}
\end{equation}
resulting in Rotta's well-known linear return-to-isotropy model,\cite{Rotta_51,Haworth_86} if $C_0=2/3(C_R-1)$, where $C_R$ is Rotta's constant. This can be obtained from the general case by assuming constant density $\mathcal{M}_{ij}=0$, isotropic $G_{ij}$ and isotropic small scales $H_{ij}=0$.

\noindent\textbf{(3) GLM in VD flows with small scale isotropy.} If $G_{ij}$ is chosen to be anisotropic, while the small scales to be isotropic, $H_{ij}\!=\!0$, \Eqre{eq:bvd} simplifies to
\begin{align}
&2k\left.\ld{b_{ij}}\right|^{\varrho,\mathrm{i}}_\mathrm{GLM} =  2k\left[G_{ik}b_{jk} + G_{jk}b_{ik} + \frac{1}{3}(G_{ij}+G_{ji})\right]\nonumber\\
&+\mathcal{M}^\mathrm{d}_{ij} - \left[2\left(\mathcal{M}+G_{kl}\rmean{v\rf_kv\rf_l}\right) + 3\phi_{\scriptscriptstyle I}\varepsilon\right] b_{ij} - \frac{2}{3}G_{kl}\rmean{v\rf_kv\rf_l}\delta_{ij},\label{eq:bvdglm}
\end{align}
where the superscript ``i'' indicates presumed small scale isotropy. This case amounts to applying the GLM in VD flows, where $G_{ij}$ is given by any anisotropic model, such as discussed by Pope.\cite{Pope_94} It can be obtained from the general case by assuming variable density $\mathcal{M}_{ij}\ne0$, anisotropic $G_{ij}$ and isotropic small scales $H_{ij}=0$.

\noindent\textbf{(4) SLM in VD flows with small scale anisotropy.} If an isotropic $G_{ij}$ is chosen to be employed in the extended model, \Eqre{eq:TLM}, for VD flows, in \Eqre{eq:bvd} the terms involving $G_{ij}$ cancel as in the constant-density case and $b_{ij}$ will be governed by
\begin{equation}
\begin{split}
\left.2k\ld{b_{ij}}\right|^{\varrho,\mathrm{a}}_\mathrm{SLM} & = \mathcal{M}^\mathrm{d}_{ij} + \varepsilon\phi_{\scriptscriptstyle D}H^\mathrm{d}_{ij}\\
&\quad -\left[2\mathcal{M} + \big(3\phi_{\scriptscriptstyle I}+\phi_{\scriptscriptstyle D}H_{kk}\big)\varepsilon\right] b_{ij},
\end{split}\label{eq:bvda}
\end{equation}
where the superscript ``a'' denotes presumed small scale anisotropy. This can be obtained from the general case by assuming variable density $\mathcal{M}_{ij}\ne0$, isotropic $G_{ij}$ and anisotropic small scales $H_{ij}\ne0$.

\noindent\textbf{(5) SLM in VD flows with small scale isotropy.} \Eqre{eq:bvda} further simplifies in a case that adheres to Kolmogorov's hypothesis of local isotropy, i.e.\ $H_{ij}=0$,
\begin{equation}
\quad\qquad\left.2k\ld{b_{ij}}\right|^{\varrho,\mathrm{i}}_\mathrm{SLM} = \mathcal{M}^\mathrm{d}_{ij} - \left(2\mathcal{M} + 3\phi_{\scriptscriptstyle I}\varepsilon\right) b_{ij}.\label{eq:bvdi}
\end{equation}
This can be obtained from the general case by assuming variable density $\mathcal{M}_{ij}\ne0$, isotropic $G_{ij}$ and isotropic small scales $H_{ij}=0$.

In constant-density high-Reynolds-number flows that are free of large strains and rotations the small scales are isotropic. In contrast, Rayleigh-Taylor flows are an\-i\-so\-trop\-ic at all times at both large and small scales.\cite{Livescu_07,Livescu_08,Livescu_09,Livescu_09c,Chung_10} These variable-density flows are at moderate Reynolds numbers and free of large distortions. To this end we are concerned with both $b_{ij}$ and $d_{ij}$.

In the homogeneous RT DNS of Livescu \& Ristorcelli,\cite{Livescu_07,Livescu_08} $\mathcal{P}_{ij}=0$ thus $\mathcal{M}_{ij}$ is the sole source of $b_{ij}$ which quickly vanishes in the initial stages of the flow evolution. On the other hand, $b_{ij}$ indicates anisotropy for the full extent of the simulations. This is most likely due to the suddenly decreasing turbulent Reynolds number closely following the behavior of $\mathcal{M}_{ij}$. As the significance of the non-linear term in the Navier-Stokes equation (compared to viscous dissipation) decreases, the efficiency of the non-linear cascade, that would transfer energy from the large to the small scales, weakens. This effectively locks in the anisotropy structure of $b_{ij}$ which thus dissipates very slowly.

The VD PDF model is designed to account for a source of anisotropy that is independent of $\mathcal{M}_{ij}$ and a (possibly) initially nonzero $b_{ij}$. That is, in principle, the model is capable of capturing the large and small scale anisotropy of variable-density, strained or low-\emph{Re} flows through the combined effects of $G_{ij}$ and $H_{ij}$. For generality, the following discussion supposes an initially zero $b_{ij}$.

The special cases of \Eqre{eq:bvd} above show that there are two independent sources of anisotropy in the VD velocity field, represented by the SDE \Eqrs{eq:TLM} (without significant shear-production): via $G_{ij}$ and via $H_{ij}$. This is indicated by
\begin{enumerate}
\item \Eqre{eq:bvdglm}, where $H_{ij}\!=\!0$ and the source is proportional to the anisotropic $G_{ij}$, (large scale anisotropy), and,
\item \Eqre{eq:bvda}, where the isotropic $G_{ij}$ has no effect and the source is proportional to $H_{ij}$, (small scale anisotropy).
\end{enumerate}
In the absence of shear production, $\mathcal{P}_{ij}=0$, embedded in the Lagrangian derivative, \Eqre{eq:bvdi} prescribes a decay of $b_{ij}$ after $\mathcal{M}_{ij}$ vanishes. If we assume $H_{ij}$ to be nonzero, we can take two routes in defining $G_{ij}$:
\begin{enumerate}
\item \emph{Isotropic $G_{ij}$, special case (4).} In this case $b_{ij}$ is governed by \Eqre{eq:bvda}. If one chooses an isotropic $G_{ij}$ in VD flows, \Eqre{eq:bvda} shows that a nonzero $H_{ij}$ can sustain $b_{ij}$ independently even after $\mathcal{M}_{ij}$ vanished, c.f.\ \Eqre{eq:br}. An isotropic $G_{ij}$ also means that the effects of large density fluctuations on the mean velocity are assumed to be isotropic, since according to \Eqre{eq:mm}, $-\rmean{v\rf p\rf\!,_i} + \mu\rmean{v\rf\nabla^2v_i\rf} = -\rmean{v_i\ff}/T$, where $T$ denotes some function of a time scale, e.g.\ $k/\varepsilon$.
\item \emph{Anisotropic $G_{ij}$, the general case.} In this case $b_{ij}$ is governed by \Eqre{eq:bvd}. The choice of an anisotropic $G_{ij}$ combined with a nonzero $H_{ij}$ opens up all the possibilities of the SDE \Eqrs{eq:TLM} to correctly capture $b_{ij}$ in VD flows. This most general formulation seems to be capable of representing nonzero $b_{ij}$ patterns that are more complex than that of a homogeneous RT flow.
\end{enumerate}

In summary, we showed that the VD SDE \Eqrs{eq:TLM} allows for two independent sources of anisotropy, proportional to $G_{ij}$ and $H_{ij}$, respectively. The question of whether $G_{ij}$, which is capable of sustaining anisotropy in the absence of shear production, should be assumed isotropic or anisotropic in general VD flows remains an open question. Functional forms of $G_{ij}$ and $H_{ij}$ are the subject of the next paper, Ref.\ \onlinecite{Bakosi_10c}. Their combined effect should
\begin{enumerate}
\item Create the correct $b_{ij}$ structure in VD flows by modeling the processes in the Reynolds stress budget as \Eqre{eq:rsmodel} prescribes,
\item $G_{ij}$ should provide the correct representation of the processes $-\rmean{v\rf p\rf\!,_i} + \mu\rmean{v\rf\nabla^2v_i\rf} = G_{ij}\rmean{v_j\ff}$ in the mean velocity equation \Eqrs{eq:mmv}, and,
\item $G_{ij}$ and $H_{ij}$ should provide the correct representation of the variable-density effects on the turbulent kinetic energy as given by \Eqre{eq:cGCH}.
\end{enumerate}

\subsection{The functional forms of $\phi_{\scriptscriptstyle I}$ and $\phi_{\scriptscriptstyle D}$}
This section establishes further constraints on the functions $\phi_{\scriptscriptstyle I}$ and $\phi_{\scriptscriptstyle D}$, governing the diffusion terms in the VD SDE \Eqrs{eq:TLM},
\begin{equation}
\left(\phi_{\scriptscriptstyle I}\varepsilon\right)^{1/2}\mathrm{d}W_i + \left(\phi_{\scriptscriptstyle D}\varepsilon\right)^{1/2}h_{ij}\mathrm{d}W'_j.\label{eq:diffusion}
\end{equation}

As a reminder, the mathematical representation of a diffusion process requires $\phi_{\scriptscriptstyle I}$, $\phi_{\scriptscriptstyle D}$ and $h_{ij}$ to be non-negative and bounded, which also ensures realizability. From the modeling point of view, it is customary to construct these functions so that only the isotropic part of \Eqre{eq:diffusion} affects the turbulent kinetic energy budget, while only the deviatoric part affects the $b_{ij}$ budget. This is ensured if
\begin{equation}
\phi_{\scriptscriptstyle I} = C_0 - \frac{2}{3}H\phi_{\scriptscriptstyle D},\label{eq:phiI}
\end{equation}
which can be easily seen from the corresponding terms in the $k$ equation \Eqrs{eq:km},
\begin{equation}
\left(\frac{3}{2}\phi_{\scriptscriptstyle I} + \phi_{\scriptscriptstyle D}H\right)\varepsilon = \frac{3}{2}C_0\varepsilon,\label{eq:diffk}
\end{equation}
and the $b_{ij}$ equation \Eqrs{eq:bvd},
\begin{equation}
\big(3\phi_{\scriptscriptstyle I} + 2\phi_{\scriptscriptstyle D}H\big)\varepsilon b_{ij} = 3C_0\varepsilon b_{ij}.\label{eq:diffb}
\end{equation}
\Eqres{eq:diffk} and \Eqrs{eq:diffb} respectively show that with the choice of $\phi_{\scriptscriptstyle I}$ in \Eqre{eq:phiI}, $k$ will not be affected by $H_{ij}$ or $\phi_{\scriptscriptstyle_D}$, while $b_{ij}$ will only be affected by $H^\mathrm{d}_{ij}$ and $\phi_{\scriptscriptstyle D}$, see \Eqre{eq:bvd}. \Eqre{eq:phiI} ensures that the VD model reduces to its constant-density counterpart with small scale isotropy if $\mathcal{M}\to0$: \Eqre{eq:km} governing $k$ approaches \Eqre{eq:inckm} independent of $H_{ij}$.

Since $\phi_{\scriptscriptstyle I}$ must be non-negative, the following bounds on $\phi_{\scriptscriptstyle D}$ must also be imposed:
\begin{equation}
0 \le \phi_{\scriptscriptstyle D} \le \frac{3C_0}{2H}.\label{eq:phiD}
\end{equation}
As $C_0$ and $H_{ij}$ are bounded, \Eqres{eq:phiI} and \Eqrs{eq:phiD} ensure that $\phi_{\scriptscriptstyle I}$ and $\phi_{\scriptscriptstyle D}$ are also bounded. Thus we specify $\phi_{\scriptscriptstyle D}$ as
\begin{equation}
\phi_{\scriptscriptstyle D} = [1-g(\theta)]\frac{3C_0}{2H} \qquad \textrm{with} \qquad 0 \le g(\theta) \le 1,\label{eq:phiDspec}
\end{equation}
where $g(\theta)$ is some function of the mixing state, $\theta$, with $g(\theta)\!=\!0$ in the unmixed state and $g(\theta)\!=\!1$ in the fully mixed state. The simplest specifications for $g(\theta)$ and $\theta$ are through the first two density moments\cite{Youngs_91,Linden_94,Ristorcelli_Clark_04,Livescu_08}
\begin{equation}
g(\theta) = \theta = 1 - \frac{\rv{\varrho}}{\rmean{\varrho}(1-\rmean{\varrho})}.\label{eq:g}
\end{equation}
Since $\rv{\varrho}$ is maximum in the unmixed state and vanishes in the fully mixed state, \Eqre{eq:g} ensures the required behavior of $g(\theta)$ in the extreme states.

Incorporating the new specifications in \Eqres{eq:phiI} and \Eqrs{eq:phiDspec} for $\phi_{\scriptscriptstyle I}$ and $\phi_{\scriptscriptstyle D}$ into \Eqre{eq:TLM}, the final form of the VD SDE is
\begin{align}
\mathrm{d}v^*_i & = \left(g_i - \rmean{p},_i/\varrho^* + \mu/\varrho^*\nabla^2\rmean{v}_i\right)\mathrm{d}t + G_{ij}\left(v^*_j-\fmean{v_j}\right)\mathrm{d}t\nonumber\\
& + \left[C_0g(\theta)\varepsilon\right]^{1/2}\mathrm{d}W_i + \left\{\frac{3C_0}{2H}[1-g(\theta)]\varepsilon\right\}^{1/2}\!h_{ij}\mathrm{d}W'_j,\label{eq:TLMf}
\end{align}
where $H_{ij}\!=\!h_{ik}h_{kj}$ and $H\!=\!H_{ii}/2$. \Eqre{eq:TLMf} shows that $H_{ij}$ can have any unit, as it only enters in a normalized fashion and that the only restrictions on its functional form are being symmetric non-negative semi-definite and bounded so that statistics of the process exist and are well defined. Substituting \Eqres{eq:phiI} and \Eqrs{eq:phiDspec} into \Eqres{eq:mrs} and \Eqrs{eq:km} gives the final governing equations for the Reynolds stress and turbulent kinetic energy, respectively,
\begin{align}
\begin{split}
\left.\ld{\rmean{v_i\rf v_j\rf}}\right|_\mathrm{SDE} & = \mathcal{M}_{ij} + G_{ik}\rmean{v_j\rf v_k\rf} + G_{jk}\rmean{v_i\rf v_k\rf}\\
& + C_0\varepsilon\left\{\delta_{ij} + 3[1-g(\theta)]\bigg(\frac{H_{ij}}{2H}-\frac{1}{3}\delta_{ij}\bigg)\right\},\label{eq:mrsf}
\end{split}\\
\left.\ld{k}\right|_\mathrm{SDE} & = \mathcal{M} + G_{ij}\rmean{v\rf_iv\rf_j} + \frac{3}{2}C_0\varepsilon.\label{eq:kmf}
\end{align}
These equations show that only the normalized deviatoric part of $H_{ij}$ affects the second moments and only their anisotropic part, i.e.\ $b_{ij}$. As mixing progresses $g(\theta)$ takes on values from $0$ to $1$ and diminishes the small scale anisotropy, independent of $H_{ij}$.

The developed constraints on the diffusion coefficients $\phi_{\scriptscriptstyle I}$ and $\phi_{\scriptscriptstyle D}$, \Eqres{eq:phiI} and \Eqrs{eq:phiDspec}, ensure that (1) only the isotropic part of the sum of the two stochastic diffusion terms affect the evolution of $k$ and (2) $b_{ij}$ is affected by only the normalized deviatoric part. Enforcing mathematical consistency on the constraints, a function of a mix metric, $g(\theta)$, naturally appears which ensures that the small scale anisotropy is diminished in the fully mixed state. The SDE \Eqrs{eq:TLMf} reduces to its constant-density counterpart, \Eqre{eq:GLM}, as $\varrho\rf\to0$, i.e.\ $\varrho^*\to\varrho_0$ and $g(\theta)\to1$.

\subsection{Summary of the ensemble moment equations}
We derived and analyzed the first two moment equations governed by the SDE \Eqrs{eq:TLM}. The findings can be summarized as follows:
\begin{itemize}
\item The drift, $G_{ij}(v^*_j-\fmean{v_j})$, affects the Reynolds mean velocity in VD flows. This must be taken into account by the specification of $G_{ij}$ as indicated by \Eqre{eq:mm}.
\item In contrast to moment closures, the important source of turbulence in VD flows, $\mathcal{M}_{ij}$ defined by \Eqre{eq:M}, appears closed in the PDF formulation and thus requires no explicit modeling. The relation \Eqrs{eq:rsmodel} guides the modeling of the physical processes of redistribution, transport, dissipation and the triple correlations.
\item We derived a model constraint in inhomogeneous variable-density flows with small scale anisotropy, \Eqre{eq:cGCH}, for transport and scalar effects of variable density.
\item The VD SDE \Eqrs{eq:TLM} allows for two independent sources of anisotropy, proportional to $G_{ij}$ and $H_{ij}$, respectively. $G_{ij}$ is responsible for anisotropy at the large scales, while $H_{ij}$ at the small scales. This allows the equation to capture both large ($b_{ij}$) and small ($d_{ij}$) scale anisotropy of variable-density, strained or low-\emph{Re} flows.
\item We developed constraints on the diffusion coefficients, \Eqres{eq:phiI} and \Eqrs{eq:phiDspec}. These ensure that the isotropic part of the stochastic diffusion terms in \Eqre{eq:TLM} only affects the turbulent kinetic energy, while the Reynolds stress anisotropy is only affected by the deviatoric part. A mix metric ensures vanishing small scale anisotropy of different-density species in the fully mixed state.
\item The final form of the VD SDE \Eqrs{eq:TLMf} (incorporating the constraints on the diffusion terms) reduces to the GLM for constant-density flows as $\varrho\rf\to0$.
\end{itemize}

\section{Favre moment equations}
\label{sec:Favremoments}
Variable-density flows are traditionally investigated (and modeled) using Favre-averaged variables. This section derives the Favre moment equations of the VD velocity model, \Eqre{eq:TLM}, and contrasts them to their counterparts based on the Navier-Stokes equation. This is useful in comparing the PDF formulation to existing Favre moment closures and provides insight for modeling unclosed terms.

\subsection{The MDF transport equation}
This section gives the transport equation from which the equations governing the Favre moments are derived.

The transport equation governing the joint \emph{mass density function} (MDF),
\begin{equation}
\mathscr{F}(\varrho,\bv{v};\bv{x},t) \equiv \varrho f(\varrho,\bv{v};\bv{x},t),
\end{equation}
can be obtained by multiplying \Eqre{eq:vFP} by $\varrho$ and using the law of mass conservation,
\begin{align}
\frac{\partial\mathscr{F}}{\partial t} + \frac{\partial v_i\mathscr{F}}{\partial x_i} & = -\frac{\partial}{\partial v_i}\Big[\left(g_i - \rmean{p},_i/\varrho +\mu/\varrho\nabla^2\rmean{v}_i\right)\mathscr{F}\Big]\nonumber\\
&\quad - \frac{\partial}{\partial v_i}\Big[G_{ij}\big(v_j - \fmean{v_j}\big)\mathscr{F}\Big]\nonumber\\
&\quad + \frac{1}{2}\frac{\partial^2}{\partial v_i\partial v_j}\Big[\big(\phi_{\scriptscriptstyle I}\delta_{ij}+\phi_{\scriptscriptstyle D}H_{ij}\big)\varepsilon\mathscr{F}\Big]\nonumber\\
&\quad + \textrm{density model terms,}\label{eq:FP}
\end{align}
where for simplicity, we kept $\phi_{\scriptscriptstyle I}$ and $\phi_{\scriptscriptstyle D}$ in the diffusion terms. This equation is integrated to obtain Favre moment equations. In Favre averaging the density acts as a weight, therefore the precise functional form of the density equation is unimportant from the viewpoint of deriving Favre moment equations of the velocity field. Two important characteristics of $\mathscr{F}$ are:
\begin{align}
\iint \mathscr{F}(\varrho,\bv{v};\bv{x},t)\mathrm{d}\varrho\mathrm{d}\bv{v} & = \rmean{\varrho}(\bv{x},t),\\
\iint Q(\bv{v})\mathscr{F}(\varrho,\bv{v};\bv{x},t)\mathrm{d}\varrho\mathrm{d}\bv{v} & = \rmean{\varrho}(\bv{x},t)\fmean{Q(\bv{x},t)},
\end{align}
where $Q(\bv{v})$ is almost any function\cite{Pope_85} with its Favre average $\fmean{Q}$. These attributes of $\mathscr{F}$ serve as the basis of obtaining the Favre moment equations.

\subsection{The Favre mean velocity equation: $\fmean{v_i}$}
Multiplying \Eqre{eq:FP} by the sample space variable $v_k$ and integrating each term\cite{Pope_85} yields the equation governing the modeled Favre mean velocity, $\fmean{v_i}$, governed by the SDE \Eqrs{eq:TLM}, as
\begin{equation}
\left.\frac{\partial\rmean{\varrho}\fmean{v_i}}{\partial t}\right|_\mathrm{SDE} + \frac{\partial\rmean{\varrho}\fmean{v_i}\fmean{v_j}}{\partial x_j} + \frac{\partial\rmean{\varrho}\fmean{v\ff_iv\ff_j}}{\partial x_j} = \rmean{\varrho}g_i - \rmean{p},_i + \mu\nabla^2\rmean{v}_i.\label{eq:fmm}
\end{equation}
In moment closures $\fmean{v_i}$ is typically written as $\tilde{U}_i$ or $\tilde{V}_i$. Contrasting \Eqre{eq:fmm} with its counterpart, derived from the variable-density Navier-Stokes equation \Eqrs{eq:vns},
\begin{equation}
\left.\frac{\partial\rmean{\varrho}\fmean{v_i}}{\partial t}\right|_\mathrm{NS} + \frac{\partial\rmean{\varrho}\fmean{v_i}\fmean{v_j}}{\partial x_j} + \frac{\partial\rmean{\varrho}\fmean{v\ff_iv\ff_j}}{\partial x_j} = \rmean{\varrho}g_i - \rmean{p},_i + \mu\nabla^2\rmean{v}_i,\label{eq:fme}
\end{equation}
shows that the SDE \Eqrs{eq:TLM} is consistent with the Navier-Stokes Favre mean. All terms are the same, in closed form, containing no explicit model terms.

For comparison and later use, the Reynolds-averaged mean velocity equation, derived from the Navier-Stokes \Eqre{eq:vns}, is given by \Eqre{eq:emv}, or equivalently,
\begin{equation}
\begin{split}
\left.\frac{\partial\rmean{v}_i}{\partial t}\right|_\mathrm{NS} & + \rmean{v}_j\!\cdot\!\rmean{v}_{i,j} + \rmean{v\rf_jv\rf_{i,j}} =\\
& = g_i - \rmean{v}\!\cdot\!\rmean{p},_i + \mu\rmean{v}\!\cdot\!\nabla^2\rmean{v}_i - \rmean{v\rf p\rf\!,_i} + \mu\rmean{v\rf\nabla^2v_i\rf},
\end{split}\label{eq:rmex}
\end{equation}
where the effects of the fluctuating specific volume (last two terms) are exposed and have been shown to be modeled as $-\rmean{v\rf p\rf\!,_i} + \mu\rmean{v\rf\nabla^2v_i\rf} = G_{ij}\rmean{v_j\ff}$.

The above development shows that the PDF model \Eqre{eq:TLM} is consistent with the Navier-Stokes equation in both Favre and Reynolds-averaged frameworks in the mean. The effects of large density fluctuations appear hidden in the former and explicitly modeled in the latter. This is in contrast to Favre moment closures, where they are not accounted for if a constant-density model is used to represent the Favre Reynolds stress, $\fmean{v\ff_iv\ff_j}$.

\subsection{The mass flux equation: $a_i=\rmean{\varrho\rf v_i\rf}/\rmean{\varrho}$}
The equations governing the mass flux can be obtained from $a_i=\fmean{v_i}-\rmean{v}_i$, using Eqs.\ \Eqrs{eq:fmm} and \Eqrs{eq:mmv} for the model and Eqs.\ \Eqrs{eq:fme} and \Eqrs{eq:emv} from Navier-Stokes:
\begin{align}
\left.\rmean{\varrho}\ld{a_i}\right|_\mathrm{SDE} & = -\rmean{\varrho\rf v\rf}\left(\rmean{p},_i - \mu\nabla^2\rmean{v}_i\right) + \rmean{\varrho}G_{ij}a_j,\label{eq:lmfm}\\
\left.\rmean{\varrho}\ld{a_i}\right|_\mathrm{NS} & = -\rmean{\varrho\rf v\rf}\left(\rmean{p},_i - \mu\nabla^2\rmean{v}_i\right) + \rmean{\varrho}\big(\rmean{v\rf p\rf\!,_i} - \mu\rmean{v\rf\nabla^2v_i\rf}\big).\label{eq:lmfe}
\end{align}
Comparing Eqs.\ \Eqrs{eq:lmfm} and \Eqrs{eq:lmfe} yields
\begin{equation}
-\rmean{v\rf p\rf\!,_i} + \mu\rmean{v\rf\nabla^2v_i\rf} = -G_{ij}a_j,\label{eq:amm}
\end{equation}
the same as \Eqre{eq:mm}, already obtained from contrasting the Reynolds mean velocity equations. \Eqre{eq:amm} establishes the consistent modeling of the mass flux by the PDF model and highlights the importance of $G_{ij}a_j$ as a model for the specific-volume-pressure-gradient and the specific-volume-viscous-force covariances.

The above development can be put in context with VD Favre moment closures in the Eulerian framework by expanding the Lagrangian derivative in \Eqre{eq:lmfe} to yield the mass flux equation\cite{Livescu_09}
\begin{align}
&\!\!\!\!\!\left.\rmean{\varrho}\ld{a_i}\right|_\mathrm{NS} = \frac{\partial\rmean{\varrho}a_i}{\partial t} + \frac{\partial\rmean{\varrho}\fmean{v_j}a_i}{\partial x_j} + \rmean{\varrho}a_j\big(\fmean{v_i}-a_i\big),_j + \rmean{\varrho}\!\cdot\!\rmean{v\rf_i\ivdrf} \nonumber\\
&\qquad - \rmean{\varrho}(a_ia_j),_j - \frac{\rmean{\varrho},_j}{\rmean{\varrho}}\big(\rmean{\varrho\rf v\rf_iv\rf_j}-\rmean{\varrho}\fmean{v\ff_iv\ff_j}\big) + \frac{\partial\rmean{\varrho\rf v\rf_iv\rf_j}}{\partial x_j} = \nonumber\\
&\qquad = -\rmean{\varrho\rf v\rf}\left(\rmean{p},_i - \mu\nabla^2\rmean{v}_i\right) + \rmean{\varrho}\big(\rmean{v\rf p\rf\!,_i} - \mu\rmean{v\rf\nabla^2v_i\rf}\big).
\label{eq:dadt}
\end{align}
In moment closures,\cite{Besnard_92,Gregoire_05} equations are solved for $\rmean{\varrho}$, $\fmean{v_i}$, $\fmean{v\ff_iv\ff_j}$, $\Hat{b}\!=\!-\rmean{\varrho\rf v\rf}$ (or $\rv{\varrho}$) and $a_i$. In \Eqre{eq:dadt} the effects of the unclosed terms, $\rmean{v\rf_i\ivdrf}$, $\rmean{\varrho},_j\!\rmean{\varrho\rf v\rf_iv\rf_j}$, $(\rmean{\varrho\rf v\rf_iv\rf_j}),_j$ and $\rmean{v\rf p\rf\!,_i} - \mu\rmean{v\rf\nabla^2v_i\rf}$ are modeled. In contrast, in the PDF method, \Eqre{eq:amm} shows that only the non-Boussinesq effects, $\rmean{v\rf p\rf\!,_i} - \mu\rmean{v\rf\nabla^2v_i\rf}$, require closure assumptions.

It is useful to investigate the model mass flux and mean velocity equations in the limit $\rmean{v}_i=0$. This holds exactly at arbitrary Atwood numbers in a homogeneous RT layer and at low Atwood numbers in the inhomogeneous RT layer. Also, $\rmean{v}_i\approx0$ is still a good approximation in high-Atwood-number inhomogeneous RT layers.\cite{Livescu_07} If $\rmean{v}_i=0$, then $a_i=\fmean{v_i}$, thus from \Eqres{eq:fmm} and \Eqrs{eq:lmfm} we have
\begin{equation}
g_i - \rmean{v}\!\cdot\!\rmean{p},_i - (\rmean{v\rf_jv\rf_i}),_j + \rmean{v\rf_i\ivdrf} = G_{ij}a_j.\label{eq:amm2}
\end{equation}
The same is obtained by setting $\rmean{v}_i=0$ in \Eqre{eq:mmv} with the advection term expanded as in \Eqre{eq:rmex}. While \Eqre{eq:amm} holds in inhomogeneous VD flows with large density fluctuations, \Eqre{eq:amm2} can be considered as an approximation. The importance of \Eqre{eq:amm2} is that its processes may be easier to model than those in \Eqre{eq:amm}.

\subsection{The Favre Reynolds stress equation: $\fmean{v\ff_iv\ff_j}$}
Multiplying \Eqre{eq:FP} by $(v_k-\fmean{v_k})(v_l-\fmean{v_l})$ then integrating produces the model equation governing the Favre Reynolds stress, $\fmean{v\ff_iv\ff_j}$, of the SDE \Eqrs{eq:TLM} as
\begin{equation}
\begin{split}
&\left.\frac{\partial\rmean{\varrho}\fmean{v\ff_iv\ff_j}}{\partial t}\right|_\mathrm{SDE} + \frac{\partial\rmean{\varrho}\fmean{v_k}\fmean{v\ff_iv\ff_j}}{\partial x_k} + \frac{\partial\rmean{\varrho}\fmean{v\ff_iv\ff_jv\ff_k}}{\partial x_k}\\
&\quad + \rmean{\varrho}\fmean{v\ff_iv\ff_k}\fmean{v_j}\!,_k + \rmean{\varrho}\fmean{v\ff_jv\ff_k}\fmean{v_i}\!,_k =\\
&\quad = a_i\rmean{p},_j + a_j\rmean{p},_i - \mu\big(a_i\nabla^2\rmean{v}_j + a_j\nabla^2\rmean{v}_i\big)\\
&\qquad + \rmean{\varrho}G_{ik}\fmean{v\ff_jv\ff_k} + \rmean{\varrho}G_{jk}\fmean{v\ff_iv\ff_k} + \rmean{\varrho}\big(\phi_{\scriptscriptstyle I}\delta_{ij} + \phi_{\scriptscriptstyle D}H_{ij}\big)\varepsilon.
\end{split}\label{eq:frsm}
\end{equation}
The Favre Reynolds stress equation, derived from the VD Navier-Stokes \Eqre{eq:vns}, is
\begin{equation}
\begin{split}
&\left.\frac{\partial\rmean{\varrho}\fmean{v\ff_iv\ff_j}}{\partial t}\right|_\mathrm{NS} + \frac{\partial\rmean{\varrho}\fmean{v_k}\fmean{v\ff_iv\ff_j}}{\partial x_k} + \frac{\partial\rmean{\varrho}\fmean{v\ff_iv\ff_jv\ff_k}}{\partial x_k}\\
&\quad + \rmean{\varrho}\fmean{v\ff_iv\ff_k}\fmean{v_j}\!,_k + \rmean{\varrho}\fmean{v\ff_jv\ff_k}\fmean{v_i}\!,_k =\\
&\quad = a_i\rmean{p},_j + a_j\rmean{p},_i - \mu\big(a_i\nabla^2\rmean{v}_j + a_j\nabla^2\rmean{v}_i\big)\\
&\qquad + (\mathcal{R}_{ij} - \mathcal{T}_{ij} - \varepsilon_{ij})/\rmean{v}.
\end{split}\label{eq:frse}
\end{equation}
Comparing \Eqres{eq:frsm} and \Eqrs{eq:frse} shows that the first three lines are the same. The terms proportional to $a_i$ represent the source of turbulence in the Favre-averaged framework whose role played by $\mathcal{M}_{ij}$ in the ensemble-averaged framework, c.f.\ \Eqres{eq:ersd} and \Eqrs{eq:frse} and see \Eqre{eq:Mrho} that details the effects of the fluctuating specific volume. The correspondence of the remaining terms produces the modeling constraint
\begin{equation}
\begin{split}
&\negthickspace\negthickspace(\mathcal{R}_{ij} - \mathcal{T}_{ij} - \varepsilon_{ij})/(\rmean{\varrho}\cdot\rmean{v}) =\\
& = G_{ik}\fmean{v\ff_jv\ff_k} + G_{jk}\fmean{v\ff_iv\ff_k} + \big(\phi_{\scriptscriptstyle I}\delta_{ij} + \phi_{\scriptscriptstyle D}H_{ij}\big)\varepsilon,
\end{split}\label{eq:fmr}
\end{equation}
which can be contrasted to the one obtained from the comparison of the ensemble Reynolds stress equations, \Eqre{eq:rsmodel}, to yield an expression for the triple correlations in $\mathcal{V}_{ij}$ as
\begin{align}
\mathcal{V}_{ij} & \equiv -\rmean{v\rf(v_i\rf p\rf\!,_j + v_j\rf p\rf\!,_i)} + \mu\rmean{v\rf(v_i\rf\nabla^2v_j\rf + v_j\rf\nabla^2v_i\rf)} = \nonumber\\
& = \rmean{\varrho\rf v\rf}\left[G_{ik}\rmean{v\rf_jv\rf_k} + G_{jk}\rmean{v\rf_iv\rf_k} + \big(\phi_{\scriptscriptstyle I}\delta_{ij} + \phi_{\scriptscriptstyle D}H_{ij}\big)\varepsilon\right]\nonumber\\
&\quad + \rmean{\varrho}\!\cdot\!\rmean{v}\left(G_{ik}a_ja_k + G_{jk}a_ia_k\right).\label{eq:Vmodel}
\end{align}
Similar to the ensemble-averaged framework, the half of the trace of \Eqre{eq:fmr} produces
\begin{equation}
(\mathcal{R} - \mathcal{T} - \varepsilon)/(\rmean{\varrho}\cdot\rmean{v}) = G_{ij}\fmean{v\ff_iv\ff_j} + \left(\frac{3}{2}\phi_{\scriptscriptstyle I} + \phi_{\scriptscriptstyle D}H\right)\varepsilon.\label{eq:fmk}
\end{equation}

A set of modeling constraints, based on the Favre Reynolds stress and turbulent kinetic energy, \Eqres{eq:fmr} and \Eqrs{eq:fmk}, has been derived. These are analogous to \Eqres{eq:rsmodel} and \Eqrs{eq:cGCH} in the ensemble framework and do not explicitly involve the triple correlations, $\mathcal{V}_{ij}$.

\subsection{Summary of the Favre moment equations}
We analyzed the Favre velocity moment equations, derived from the PDF model, that require modeling in VD moment closures. The main findings of this section can be summarized as follows:
\begin{itemize}
\item The PDF and Navier-Stokes Favre mean velocity equations are consistent. The effects of the fluctuating specific volume, hidden in the Favre formulation, are explicitly modeled in the ensemble framework, discussed in Sec.\ \ref{sec:vPDF-moments}.
\item We derived a modeling constraint that relates the model tensor $G_{ij}$ to the effects of the fluctuating specific volume, \Eqre{eq:amm2}, which is as an approximation for \Eqre{eq:amm} for flows with $\rmean{v}_i\approx0$.
\item We showed that the source of turbulence in VD flows in the Favre-averaged framework (the effect of the mass flux, $a_i$) appears closed in the PDF formulation. Pressure redistribution, transport and dissipation of the Favre Reynolds stress tensor and turbulent kinetic energy are jointly modeled according to \Eqres{eq:fmr} and \Eqrs{eq:fmk}, respectively. These two equations can be used as a guideline for specifying $G_{ij}$ and $H_{ij}$ if the Favre framework is preferred.
\end{itemize}

\section{The Boussinesq limit}
\label{sec:BoussinesqLimit}
The well-known Boussinesq approximation, in which the effects of variable density are only retained in the body force in the Navier-Stokes equation, is used in many situations. It is necessary to ensure that the VD SDE \Eqrs{eq:TLM} can account for the Boussinesq limit.

\subsection{The instantaneous Navier-Stokes equation: $v_i$}
To introduce the notation the VD and Boussinesq Navier-Stokes equations are given.

We start from two separate forms, written for the variable-density case and the Boussinesq case, respectively,
\begin{align}
\left.\varrho\ld{v_i}\right|_\mathrm{VD} & = \varrho g_i - p,_i + \mu\nabla^2v_i,\label{eq:vd}\\
\left.\varrho_0\ld{v_i}\right|_\mathrm{B} & = \varrho g_i - p,_i + \mu\nabla^2v_i,
\end{align}
where $\ild{(\cdot)}$ denotes the derivative along an instantaneous Lagrangian path. The density on the left hand side of \Eqre{eq:vd}, with the aid of the specific volume, $\varrho=1/v$, can be decomposed as $\varrho=1/(\rmean{v}+\xi v\rf)$. Thus the above two equations can be written in a combined form that explicitly shows the departure from the Boussinesq-limit:
\begin{equation}
\ld{v_i} = \varrho(\rmean{v}+\xi v\rf)g_i - (\rmean{v}+\xi v\rf)p,_i + \mu(\rmean{v}+\xi v\rf)\nabla^2v_i.\label{eq:vnsB}
\end{equation}
The non-dimensional ordering parameter, $\xi\!=\![0,1]$, distinguishes between the Boussinesq case with $\xi\!=\!0$, when $\varrho=\varrho_0=\rmean{\varrho}=1/\rmean{v}$, corresponding to
\begin{equation}
\left.\ld{v_i}\right|_\mathrm{B} = \varrho\rmean{v}g_i - \rmean{v}p,_i + \mu\rmean{v}\!\cdot\!\nabla^2v_i,\label{eq:vnsB0}
\end{equation}
and the variable-density case with $\xi\!=\!1$, when $\rmean{\varrho}\cdot\rmean{v} + \rmean{\varrho\rf v\rf} = 1$, corresponding to
\begin{equation}
\left.\ld{v_i}\right|_\mathrm{VD} = g_i - vp,_i + \mu v\nabla^2v_i,\label{eq:vnsB1}
\end{equation}
the same as \Eqre{eq:vns}.

The equations governing the first two ensemble moments of \Eqre{eq:vnsB} are investigated in the next sections. Then they will be contrasted to the VD PDF model in the Boussinesq and VD limits.

\subsection{The Navier-Stokes ensemble mean velocity equation: $\rmean{v}_i$}
The equation governing the Reynolds-averaged mean velocity, derived from \Eqre{eq:vnsB} is
\begin{equation}
\begin{split}
\ld{\rmean{v}_i} & = \rmean{\varrho}\!\cdot\!\rmean{v}g_i + \rmean{v}\!\cdot\!\rmean{p},_i + \mu\rmean{v}\!\cdot\!\nabla^2\rmean{v}_i\\
&\quad + \xi\big(\rmean{\varrho\rf v\rf}g_i - \rmean{v\rf p\rf\!,_i} + \mu\rmean{v\rf\nabla^2v_i\rf}\big).
\end{split}\label{eq:mB}
\end{equation}
Setting $\xi=0$ and $\rmean{\varrho}=1/\rmean{v}$ gives the Boussinesq limit,
\begin{equation}
\left.\ld{\rmean{v}_i}\right|_\mathrm{B} = g_i - \rmean{v}\!\cdot\!\rmean{p},_i + \mu\rmean{v}\!\cdot\!\nabla^2\rmean{v}_i,\label{eq:mB0}
\end{equation}
showing that in Boussinesq flows, where $\rmean{v}=\mathrm{const.}$, the mean velocity along an instantaneous Lagrangian path is governed by the same equation as in incompressible flows if the density is chosen as $\varrho_0=\rmean{\varrho}=1/\rmean{v}=\mathrm{const.}$,
\begin{equation}
\left.\ld{\rmean{v}_i}\right|_{\varrho_0} = g_i - \rmean{p},_i/\varrho_0 + \mu/\varrho_0\nabla^2\rmean{v}_i.\label{eq:incmv}
\end{equation}
Setting $\xi=1$ in \Eqre{eq:mB} gives the mean velocity equation in the VD case as
\begin{equation}
\left.\ld{\rmean{v}_i}\right|_\mathrm{VD} = g_i - \rmean{v}\!\cdot\!\rmean{p},_i + \mu\rmean{v}\!\cdot\!\nabla^2\rmean{v}_i - \rmean{v\rf p\rf\!,_i} + \mu\rmean{v\rf\nabla^2v_i\rf},\label{eq:mB1}
\end{equation}
the same as \Eqre{eq:emv}, since $\rmean{\varrho}\cdot\rmean{v} + \rmean{\varrho\rf v\rf} = 1$. Comparing \Eqres{eq:mB0} and \Eqrs{eq:mB1} indicates that the fluctuating specific volume is directly associated with large density fluctuations in VD flows, due to their non-negligible effect on the inertia terms and the departure from the Boussinesq limit.

\subsection{The Navier-Stokes ensemble Reynolds stress equation: $\rmean{v\rf_iv\rf_j}$}
The equation governing the ensemble-averaged Rey\-nolds stress, derived from \Eqre{eq:vnsB} is
\begin{equation}
\ld{\rmean{v\rf_iv\rf_j}} = \mathcal{M}_{ij}^\mathrm{B} + \mathcal{R}_{ij} - \mathcal{T}_{ij} - \varepsilon_{ij} + \xi\big(\mathcal{M}_{ij}^\mathrm{VD} + \mathcal{V}_{ij}\big),\label{eq:rsB}
\end{equation}
with $\mathcal{R}_{ij}$, $\mathcal{T}_{ij}$, $\varepsilon_{ij}$ and $\mathcal{V}_{ij}$ defined by Eqs.\ (\ref{eq:R}--\ref{eq:V}) and
\begin{align}
\mathcal{M}_{ij}^\mathrm{B} & = \rmean{v}\big(\rmean{\varrho\rf v\rf_j}g_i + \rmean{\varrho\rf v\rf_i}g_j\big),\label{eq:MB}\\
\begin{split}
\mathcal{M}_{ij}^\mathrm{VD} & = \rmean{\varrho}\big(\rmean{v\rf v\rf_j}g_i + \rmean{v\rf v\rf_i}g_j\big)\\
&\quad + \rmean{\varrho\rf v\rf v\rf_j}g_i + \rmean{\varrho\rf v\rf v\rf_i}g_j + \mathcal{M}_{ij},
\end{split}\\
\begin{split}
\mathcal{M}_{ij} & = \rmean{v\rf v_i\rf}\!\cdot\!\rmean{p},_j - \rmean{v\rf v_j\rf}\!\cdot\!\rmean{p},_i\\
&\quad + \mu\big(\rmean{v\rf v_i\rf}\!\cdot\!\nabla^2\rmean{v}_j + \rmean{v\rf v_j\rf}\!\cdot\!\nabla^2\rmean{v}_i\big).
\end{split}
\end{align}
As before, setting $\xi=0$ and $\rmean{\varrho}=1/\rmean{v}$ gives the Boussinesq limit,
\begin{equation}
\left.\ld{\rmean{v\rf_iv\rf_j}}\right|_\mathrm{B} = \mathcal{M}_{ij}^\mathrm{B} + \mathcal{R}_{ij} - \mathcal{T}_{ij} - \varepsilon_{ij},\label{eq:rsB0}
\end{equation}
which does not involve the mean pressure gradient but contains the buoyancy force, $g_i$, in $\mathcal{M}_{ij}^\mathrm{B}$, \Eqre{eq:MB}. Setting $\xi=1$ in \Eqre{eq:rsB} gives the ensemble VD Reynolds stress equation
\begin{equation}
\left.\ld{\rmean{v_i\rf v_j\rf}}\right|_\mathrm{VD} = \mathcal{M}_{ij} + \mathcal{R}_{ij} - \mathcal{T}_{ij} - \varepsilon_{ij} + \mathcal{V}_{ij},\label{eq:rsB1}
\end{equation}
the same as \Eqre{eq:ersd}, which does not involve the buoyancy force but contains the mean pressure gradient, $\rmean{p}_i$, as the terms proportional to $g_i$ cancel in $\mathcal{M}_{ij}^\mathrm{B}+\mathcal{M}_{ij}^\mathrm{VD}$ due to the identity $\rmean{\varrho}\!\cdot\!\rmean{v\rf v\rf_i} + \rmean{v}\!\cdot\!\rmean{\varrho\rf v\rf_i} + \rmean{\varrho\rf v\rf v\rf_i} = 0$.

\subsection{The VD Langevin model in the Boussinesq limit}
The asymptotic behavior of the SDE \Eqrs{eq:TLM} for VD flows, as it reduces to the constant-density SDE \Eqrs{eq:GLM} when $\varrho\rf\to0$, has been discussed in Sec.\ \ref{sec:vPDF-moments}. Here we investigate the question: ``Under what circumstances the moments of the VD PDF model approaches the moments of the Boussinesq limit?'' The procedure yields additional model constraints.

The equations governing the ensemble-averaged mean velocity and Reynolds stress along instantaneous Lagrangian paths, respectively, derived from the SDE \Eqrs{eq:TLM}, read
\begin{align}
\left.\ld{\rmean{v}_i}\right|_\mathrm{VD} & = g_i - \rmean{v}\!\cdot\!\rmean{p},_i + \mu\rmean{v}\!\cdot\!\nabla^2\rmean{v}_i + G_{ij}\rmean{v_j\ff},\label{eq:appmmv}\\
\begin{split}
\left.\ld{\rmean{v_i\rf v_j\rf}}\right|_\mathrm{VD} & = \mathcal{M}_{ij} + G_{ik}\rmean{v_j\rf v_k\rf} + G_{jk}\rmean{v_i\rf v_k\rf}\\
& \quad + \big(\phi_{\scriptscriptstyle I}\delta_{ij} + \phi_{\scriptscriptstyle D}H_{ij}\big)\varepsilon,\label{eq:appmrs}
\end{split}
\end{align}
the same as Eqs.\ \Eqrs{eq:mmv} and \Eqrs{eq:mrs}. These equations constitute as models for Eqs.\ \Eqrs{eq:mB} and \Eqrs{eq:rsB} in the general VD case with $\xi=1$.

\textbf{Correspondence of the mean: $\rmean{v}_i$.} Comparing \Eqres{eq:mB0} and \Eqrs{eq:appmmv} indicates that the VD PDF mean velocity equation correctly reduces to the Boussinesq Navier-Stokes mean if the following holds:
\begin{equation}
\big[G_{ij}a_j\big]_\mathrm{B} = 0.\label{eq:GB1}
\end{equation}

\textbf{Correspondence of the Reynolds stress: $\rmean{v\rf_iv\rf_j}$.} The asymptotic behavior of the model Reynolds stress equation can be most easily seen from the SDE that explicitly shows the departure from the Boussinesq limit:
\begin{align}
\mathrm{d}v^*_i & = \left(\varrho^*v^*g_i - v^*\rmean{p},_i + \mu v^*\nabla^2\rmean{v}_i\right)\mathrm{d}t + G_{ij}\left(v^*_j-\fmean{v_j}\right)\mathrm{d}t\nonumber\\
&\quad + \left(\phi_{\scriptscriptstyle I}\varepsilon\right)^{1/2}\mathrm{d}W_i + \left(\phi_{\scriptscriptstyle D}\varepsilon\right)^{1/2}h_{ij}\mathrm{d}W'_j,\label{eq:TLMB}
\end{align}
where $v^*=\rmean{v}+\xi v\rf$. This is the same as \Eqre{eq:TLM} since $v^*=1/\varrho^*$. The model Reynolds stress equation, derived from the SDE \Eqrs{eq:TLMB} becomes
\begin{equation}
\begin{split}
\left.\ld{\rmean{v_i\rf v_j\rf}}\right|_\mathrm{VD} & = \mathcal{M}_{ij}^\mathrm{B} + \xi\mathcal{M}_{ij}^\mathrm{VD} + G_{ik}\rmean{v_j\rf v_k\rf} + G_{jk}\rmean{v_i\rf v_k\rf}\\
& \quad + \big(\phi_{\scriptscriptstyle I}\delta_{ij} + \phi_{\scriptscriptstyle D}H_{ij}\big)\varepsilon.
\end{split}\label{eq:mrsB}
\end{equation}
In the Boussinesq limit, where $\xi\!=\!0$ and $\rmean{\varrho}\!=\!1/\rmean{v}$, \Eqre{eq:mrsB} as a model correctly reduces to the Boussinesq Navier-Stokes Reynolds stress \Eqre{eq:rsB0}, provided
\begin{equation}
\begin{split}
\Big[G_{ik}\rmean{v_j\rf v_k\rf} + G_{jk}\rmean{v_i\rf v_k\rf} + (\phi_{\scriptscriptstyle I}\delta_{ij} & + \phi_{\scriptscriptstyle D}H_{ij})\varepsilon\Big]_\mathrm{B} = \\
& = \mathcal{R}_{ij} - \mathcal{T}_{ij} - \varepsilon_{ij},
\end{split}
\end{equation}
i.e.\ if the effect of $\mathcal{V}_{ij}$ in the VD model vanishes. \Eqre{eq:Vmodel} shows that this can be ensured if
\begin{equation}
\big[G_{ik}a_ja_k + G_{jk}a_ia_k\big]_\mathrm{B} = 0.\label{eq:GB2}
\end{equation}
In the fully VD case with $\xi\!=\!1$ and $\rmean{\varrho}\cdot\rmean{v\rf v\rf_i} + \rmean{v}\cdot\rmean{\varrho\rf v\rf_i} + \rmean{\varrho\rf v\rf v\rf_i} = 0$, \Eqre{eq:mrsB} correctly gives the modeled Navier-Stokes Reynolds stress \Eqre{eq:rsB1}.

We obtained two additional model constraints, \Eqres{eq:GB1} and \Eqrs{eq:GB2}, to ensure the correct asymptotic behavior of the first two moments of the VD PDF model in the Boussinesq limit. The PDF model equations for $\rmean{v}_i$ and $\rmean{v\rf_iv\rf_j}$ can now be made consistent with that of the Navier-Stokes \Eqre{eq:vnsB} with $\xi\!=\!0$.

\begin{table*}[t]
\caption{\label{tab:models}
\emph{Summary of main results.} The PDF model for variable-density flows and its model constraints, involving the coefficients $G_{ij}$, $H_{ij}=h_{ik}h_{kj}$ and $g(\theta)$, related to various physical processes (defined at the right) at the first two levels of statistical moments. Note that the equation-pairs marked by ($\stackrel{\textrm{R1}}{=}$ and $\stackrel{\textrm{R2}}{=}$) and ($\stackrel{\textrm{F1}}{=}$ and $\stackrel{\textrm{F2}}{=}$) express equivalent constraints in the Reynolds and Favre-averaged frameworks, respectively.}
\hrule\addvspace{2pt}\hrule
\begin{equation*}
\mathrm{d}v^*_i = \big(g_i - \rmean{p},_i/\varrho^* + \mu/\varrho^*\nabla^2\rmean{v}_i\big)\mathrm{d}t + G_{ij}\left(v^*_j-\fmean{v_j}\right)\mathrm{d}t + \left(\phi_{\scriptscriptstyle I}\varepsilon\right)^{1/2}\mathrm{d}W_i + \left(\phi_{\scriptscriptstyle D}\varepsilon\right)^{1/2}h_{ij}\mathrm{d}W'_j
\end{equation*}\\[-0.5cm]
\begin{align*}
-\rmean{v\rf p\rf\!,_i} + \mu\rmean{v\rf\nabla^2v_i\rf} & = G_{ij}\rmean{v_j\ff} & \mathcal{R}_{ij} & = \rmean{v}\!\cdot\!\rmean{p\rf(v\rf\!_{i,j} + v\rf\!_{j,i})}\\
\mathcal{R}_{ij} - \mathcal{T}_{ij} - \varepsilon_{ij} + \mathcal{V}_{ij} & \stackrel{\textrm{R1}}{=} G_{ik}\rmean{v_j\rf v_k\rf} + G_{jk}\rmean{v_i\rf v_k\rf} + \big(\phi_{\scriptscriptstyle I}\delta_{ij} + \phi_{\scriptscriptstyle D}H_{ij}\big)\varepsilon & \mathcal{T}_{ij} & = \rmean{v}\left[(\rmean{v_i\rf p\rf}),_j + (\rmean{v_j\rf p\rf}),_i - \mu\nabla^2\rmean{v_i\rf v_j\rf}\right]\\
\mathcal{R} - \mathcal{T} + \mathcal{V} & \stackrel{\textrm{R2}}{=} G_{ij}\rmean{v\rf_iv\rf_j} + \left(1 + \textstyle\frac{3}{2}\displaystyle \phi_{\scriptscriptstyle I} + \textstyle\displaystyle \phi_{\scriptscriptstyle D}H\right)\varepsilon & \varepsilon_{ij} & = 2\mu\rmean{v}\!\cdot\!\rmean{v\rf\!_{i,k} v\rf\!_{j,k}}\\
(\mathcal{R}_{ij} - \mathcal{T}_{ij} - \varepsilon_{ij})/\rmean{v} & \stackrel{\textrm{F1}}{=} \rmean{\varrho}G_{ik}\fmean{v\ff_jv\ff_k} + \rmean{\varrho}G_{jk}\fmean{v\ff_iv\ff_k} + \rmean{\varrho}\big(\phi_{\scriptscriptstyle I}\delta_{ij} + \phi_{\scriptscriptstyle D}H_{ij}\big)\varepsilon & \mathcal{V}_{ij} & = -\rmean{v\rf(v_i\rf p\rf\!,_j + v_j\rf p\rf\!,_i)} + \mu\rmean{v\rf(v_i\rf\nabla^2v_j\rf + v_j\rf\nabla^2v_i\rf)}\Big.\\
(\mathcal{R} - \mathcal{T} - \varepsilon)/\rmean{v} & \stackrel{\textrm{F2}}{=} \rmean{\varrho}G_{ij}\fmean{v\ff_iv\ff_j} + \rmean{\varrho}\left(\textstyle\frac{3}{2}\displaystyle \phi_{\scriptscriptstyle I} + \textstyle\displaystyle \phi_{\scriptscriptstyle D}H\right)\varepsilon & \mathcal{R} & = \mathcal{R}_{ii}/2; \quad \mathcal{T} = \mathcal{T}_{ii}/2; \quad \varepsilon = \varepsilon_{ii}/2; \quad \mathcal{V} = \mathcal{V}_{ii}/2\\
\Big[G_{ij}\rmean{v_j\ff}\Big]_\mathrm{B} = 0; & \qquad \Big[G_{ik}\rmean{v_j\ff}\!\cdot\!\rmean{v_k\ff} + G_{jk}\rmean{v_i\ff}\!\cdot\!\rmean{v_k\ff}\Big]_\mathrm{B} = 0 & \phi_{\scriptscriptstyle I} & = C_0g(\theta); \quad \phi_{\scriptscriptstyle D} = \frac{3C_0}{2H}[1-g(\theta)]; \quad H=H_{ii}/2
\end{align*}
\hrule\addvspace{2pt}\hrule
\end{table*}

\subsection{Summary of constraints; VD and Boussinesq forms of the PDF model}
The set of equations, indicative of how the various physical processes are represented by the PDF model at the first two levels of statistical moments, have been summarized in Table \ref{tab:models}. This is the starting point for developing possible functional forms of the coefficients, $G_{ij}$, $H_{ij}$ and $g(\theta)$.

In the VD case the stochastic model equation is
\begin{align}
\mathrm{d}v^*_{i\:\mathrm{VD}} & = \left(g_i - v^*\rmean{p},_i + \mu v^*\nabla^2\rmean{v}_i\right)\mathrm{d}t + G_{ij}\left(v^*_j-\fmean{v_j}\right)\mathrm{d}t\nonumber\\
&\quad + \left(\phi_{\scriptscriptstyle I}\varepsilon\right)^{1/2}\mathrm{d}W_i + \left(\phi_{\scriptscriptstyle D}\varepsilon\right)^{1/2}h_{ij}\mathrm{d}W'_j.
\end{align}

Alternatively, if the development of a Boussinesq model is sought, the following equation automatically satisfies the requirements of the Boussinesq limit, Eqs.\ \Eqrs{eq:GB1} and \Eqrs{eq:GB2}:
\begin{align}
\mathrm{d}v^*_{i\:\mathrm{B}} & = \left(\varrho^*\rmean{v}g_i - \rmean{v}\!\cdot\!\rmean{p},_i + \mu\rmean{v}\!\cdot\!\nabla^2\rmean{v}_i\right)\mathrm{d}t + G_{ij}\left(v^*_j-\rmean{v}_j\right)\mathrm{d}t\nonumber\\
&\quad + \left(\phi_{\scriptscriptstyle I}\varepsilon\right)^{1/2}\mathrm{d}W_i + \left(\phi_{\scriptscriptstyle D}\varepsilon\right)^{1/2}h_{ij}\mathrm{d}W'_j,
\end{align}
whose first two moment equations can be easily seen to reduce to Eqs.\ \Eqrs{eq:mB0} and the modeled \Eqrs{eq:rsB0} with $\mathcal{V}_{ij}=0$.

\section{Conclusion}
\label{sec:vPDF-summary}
We developed and discussed fundamental ingredients necessary for extending Langevin-type models from constant-density shear flows to variable-density pressure-gradient-driven turbulence. A forthcoming article\cite{Bakosi_10c} will combine these ideas with the ones on active scalar mixing\cite{Bakosi_10} to specify a joint probability density function (PDF) model for mixing-driven variable-density (VD) hydrodynamics and will present validation for Rayleigh-Taylor flows.

In this paper we proposed a stochastic differential equation (SDE) for modeling the instantaneous velocity increments of Lagrangian particles in VD turbulence. The functional form of the equation and its consequences have been discussed in detail. Several consistency conditions and constraints, based on mathematical and physical considerations, have been developed. These results, summarized in Table \ref{tab:models}, are all rigorous mathematical consequences of the particular functional form of the proposed model equation and some basic physical considerations in VD flows. 

In particular, we showed that the widely used generalized Langevin model\cite{Haworth_86} for the velocity PDF in constant-density flows can be extended to VD flows to include the following features:
\begin{enumerate}
\item \emph{Representing variable-density effects.} The extended model incorporates the effects of large density fluctuations due to non-uniform species concentrations on the fluid particle momentum in variable-density turbulence. This includes:
  \begin{enumerate}
  \item VD effects on the inertia and mean forces terms in closed form. These are cubic (as well as quadratic) non-linearities in the Navier-Stokes equation and the mixing arising from the strong coupling between the density and velocity fields.
  \item VD effects on the fluctuating pressure gradient and viscous forces in modeled form.
  \end{enumerate}

\item \emph{Closed mass flux and density-specific-volume covariance.} In variable-density flows, these processes relate to a primary mechanism that generates turbulent kinetic energy through the mean pressure gradient. The joint PDF model for mixing-driven hydrodynamics is so constructed that the effects of the mass flux and the density-specific-volume covariance appear in closed form. Consequently, to represent these processes no explicit modeling is necessary, no additional equations need to be solved and the representation of these processes is mathematically and physically consistent. In VD moment closures these quantities require separate equations to be solved and their unclosed terms approximated.

\item \emph{Independence from the density model.} The formulation for the momentum equation is independent of the particular functional form of the density PDF model, but requires the knowledge of the instantaneous density, e.g.\ in the form of a SDE, such as discussed in Ref.\ \onlinecite{Bakosi_10}. This allows the momentum equation to be coupled to any mixing model, representing the fluid density.

\item \emph{Consistent representation of the small scale an\-i\-so\-tropy.} Turbulence anisotropy is important in predicting the correct mixing state, mixing efficiency and the production of turbulent kinetic energy. Compared to most second order moment closures that model the large scale anisotropy and assume isotropic small scales, the developed stochastic equation represents anisotropy at both the large and the small scales. This is important in non-equilibrium, highly distorted, low- or moderate-Reynolds-number or variable-density flows.

\item \emph{Reduces to the original model for constant-density shear flows.} We showed that the model coefficients can be designed so that the stochastic VD model reduces to the constant-density case in the limit of vanishing density fluctuations.
\end{enumerate}

\textbf{Construction of a PDF model.} A joint PDF model can now be constructed for variable-density turbulence where the stochastic density and velocity fields are coupled at the instantaneous level. The main ingredients for the hydrodynamics are collected in Table \ref{tab:models}, which is the essence of the paper. The VD momentum equation (first line) must be coupled to a fluid mass density PDF model, such as discussed in Ref.\ \onlinecite{Bakosi_10}, that represents conservation of mass and the mixing. Then, in the velocity model, the tensors $G_{ij}$ and $H_{ij}$, the mix metric function, $g(\theta)$, and the kinetic energy dissipation rate, $\varepsilon$, must be specified. Finally, the system must be augmented by an equation of state. The SDEs are discretized and numerically integrated in time for a large number of Lagrangian particles (representing the flow itself) in a Monte-Carlo fashion. Such a joint PDF model and its predictions for Rayleigh-Taylor flows are discussed in Ref.\ \onlinecite{Bakosi_10c}.

\section*{Acknowledgements}
J.\ Waltz and J.\ D.\ Schwarzkopf are gratefully acknowledged for a series of informative discussions. This work was performed under the auspices of the U.S.\ Department of Energy under the Advanced Simulation and Computing Program.

\appendix

\bibliographystyle{physfluids.bst}
\bibliography{jbakosi}

\end{document}